\documentclass[reprint,superscriptaddress]{revtex4-2}

\usepackage{times,graphicx,siunitx,calc,amsmath}
\usepackage[hidelinks]{hyperref}

\usepackage{float}
\makeatletter
\let\newfloat\newfloat@ltx
\makeatother
\usepackage{algorithm}

\usepackage{algorithmicx}
\usepackage{algpseudocode}

\renewcommand{\vec}[1]{\mathbf{#1}}
\newcommand{\uvec}[1]{\hat{\mathbf{#1}}}
\newcommand{\re}{\text{Re}}
\DeclareSIUnit{\Molar}{\textsc{m}} 

\begin{document}

\title{Microscopic motility of isolated \emph{E. coli} flagella} 

\author{Franky Djutanta}
\thanks{These authors contributed equally and reserve the right to change order for first authorship.}
\affiliation{Biodesign Center for Molecular Design and Biomimetics at the Biodesign Institute, Arizona State University, Tempe, AZ, 84287, USA}
\affiliation{School for Engineering of Matter, Transport and Energy, Arizona State University, Tempe, AZ, 84287, USA}

\author{Peter T. Brown}
\thanks{These authors contributed equally and reserve the right to change order for first authorship.}

\author{Bonfilio Nainggolan}

\author{Alexis Coullomb}
\affiliation{Center for Biological Physics and Department of Physics, Arizona State University, Tempe, AZ, 84287, USA}

\author{Sritharini Radhakrishnan}
\affiliation{Biodesign Center for Molecular Design and Biomimetics at the Biodesign Institute, Arizona State University, Tempe, AZ, 84287, USA}
\affiliation{School of Electrical, Computer and Energy Engineering, Arizona State University, Tempe, AZ, 84287, USA}

\author{Jason Sentosa}
\affiliation{Center for Biological Physics and Department of Physics, Arizona State University, Tempe, AZ, 84287, USA}

\author{Bernard Yurke}
\affiliation{Micron School of Materials Science and Electrical and Computer Engineering Department, Boise State University, Boise, ID, 83725, USA}

\author{Rizal F. Hariadi}
\email{rhariadi@asu.edu}
\affiliation{Biodesign Center for Molecular Design and Biomimetics at the Biodesign Institute, Arizona State University, Tempe, AZ, 84287, USA}
\affiliation{Center for Biological Physics and Department of Physics, Arizona State University, Tempe, AZ, 84287, USA}

\author{Douglas P. Shepherd}
\email{dpsheph1@asu.edu}
\affiliation{Center for Biological Physics and Department of Physics, Arizona State University, Tempe, AZ, 84287, USA}

\date{\today}


\begin{abstract}
\noindent{The fluctuation-dissipation theorem describes the intimate connection between the Brownian diffusion of thermal particles and their drag coefficients. In the simple case of  spherical particles, it takes the form of the Stokes-Einstein relationship that links the particle geometry, fluid viscosity, and diffusive behavior. However, studying the fundamental properties of microscopic asymmetric particles, such as the helical-shaped propeller used by \emph{E. coli}, has remained out of reach for experimental approaches due to the need to quantify correlated translation and rotation simultaneously with sufficient spatial and temporal resolution. To solve this outstanding problem, we generated volumetric movies of fluorophore-labeled, freely diffusing, isolated \emph{E. Coli} flagella using oblique plane microscopy. From these movies, we extracted trajectories and determined the hydrodynamic propulsion matrix directly from the diffusion of flagella via a generalized Einstein relation. Our results validate prior proposals, based on macroscopic wire helices and low Reynolds number scaling laws, that the average flagellum is a highly inefficient propeller. Specifically, we found the maximum propulsion efficiency of flagella is less than \SI{5}{\percent}. Beyond extending Brownian motion analysis to asymmetric 3D particles, our approach opens new avenues to study the propulsion matrix of particles in complex environments where direct hydrodynamic approaches are not feasible.}
\end{abstract}

\maketitle

\begin{figure*}[ht]
	\centering
	\includegraphics[width=1\linewidth]{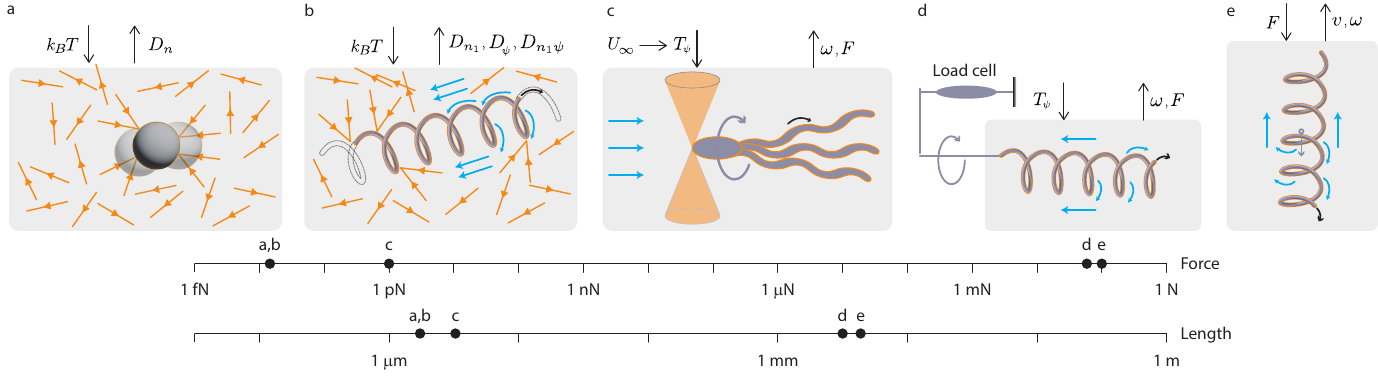}
	\caption[Evaluation of  hydrodynamics at low Reynolds number]{\textbf{Evaluation of hydrodynamics at low Reynolds number}.
	\emph{Top:}
	(\textbf{a}) An individual spherical particle is bombarded by fluid molecules (orange arrows) due to thermal fluctuations $k_B T$, resulting in Brownian motion~\cite{Perrin1909}.
	(\textbf{b}) An individual flagellum is bombarded by fluid molecules (orange arrows) due to thermal fluctuations $k_B T$, resulting in Brownian motion, with the fluid surrounding the helix mediating the interaction between translation and rotation (blue arrows).
	(\textbf{c}) A living \emph{E. coli} trapped in an optical tweezer generating a torque $T_\psi$ in response to a given fluid flow $U_\infty$, at $\re \approx 10^{-4}$, produces a bead displacement $\Delta z$ equivalent to a force value around \SI{1}{\pico\newton}~\cite{chattopadhyay2006swimming}.
	(\textbf{d}) A motor generates torque $T_\psi$ on the end of a mm-helical wire in a $\re \approx 10^{-3}$ fluid, causing the helix  to rotate at a rate of $\omega$ and to produce a force $F$ of \SI{\sim100}{\milli\newton}, as measured by a load cell~\cite{rodenborn2013propulsion}.
	(\textbf{e}) A mm-scale wire free-falls in corn syrup medium, at $\re \approx 10^{-1}$, subjected to a gravitational force $F$ of about \SI{10}{\milli\newton}.
	\emph{Bottom:} Applied force and length scales for each experimental approach.
\label{fig:figure-1}}
\end{figure*}

The original observations of Brownian motion~\cite{Brown1828,Perrin1909} and their connection to fluid drag coefficients described by Einstein's seminal paper in 1905~\cite{einstein1905erzeugung} provided a window into the deep connection between the equilibrium statistical properties and hydrodynamic response functions of a physical system. More generically, this connection is described by the fluctuation-dissipation theorem (FDT)~\cite{kubo1966fluctuation}. For a diffusing sphere, the FDT relates the experimentally accessible statistical signature of unconfined Brownian motion, the linear time dependence of the mean squared displacement (MSD), with the fluid drag coefficient and the fluid viscosity through the Stokes-Einstein relation. A common application of the Stokes-Einstein relation is the characterization of the size distribution, viscosity ($\eta$), and physical confinement of spherical particles at low Reynolds number ($\re \ll 1$) (Fig.~\ref{fig:figure-1}a). In these experiments, single particle tracking or spatio-temporal correlation functions characterize the motion of particles dispersed into a fluid by measuring the diffusion coefficient, D.

\begin{figure*}[ht]
	\centering
	\includegraphics[width=0.7\linewidth]{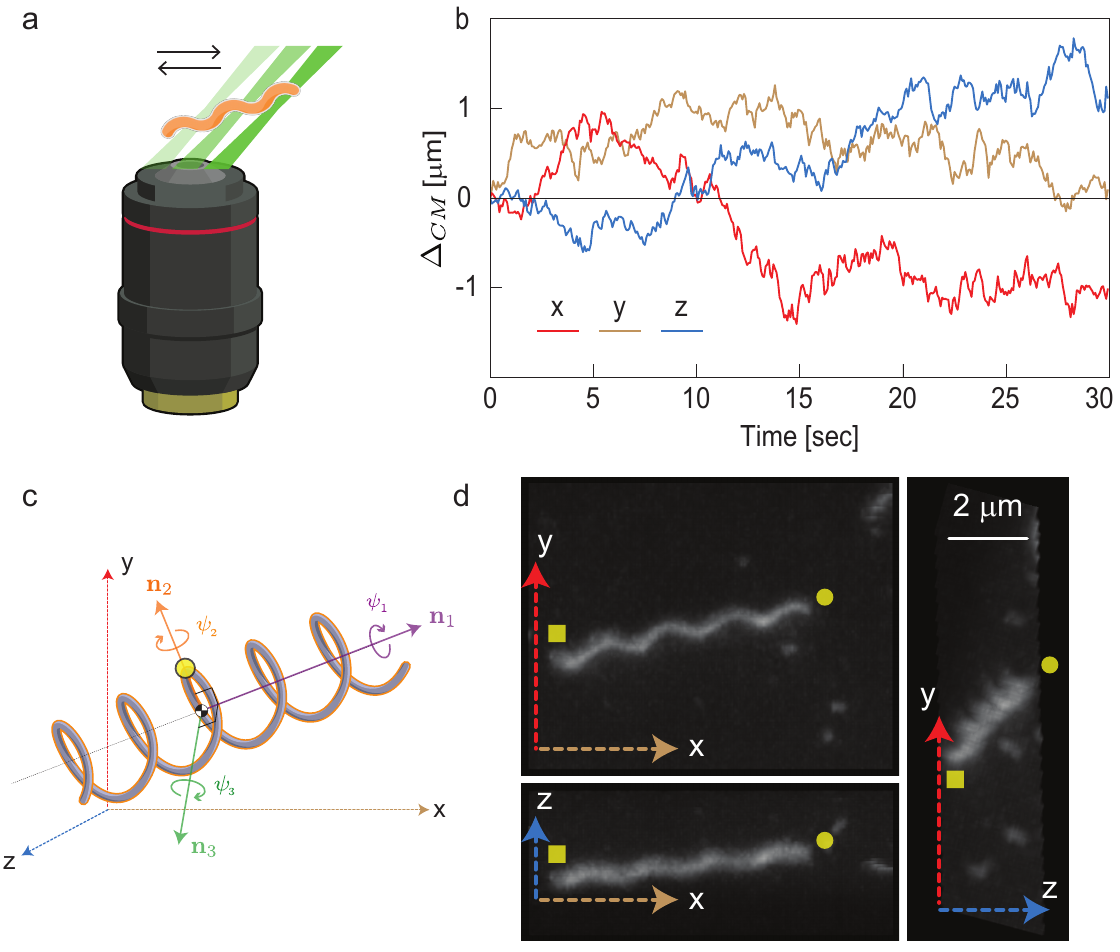}
    \caption{
    \textbf{3D tracking of diffusing flagella using an oblique plane light sheet microscope}.
    (\textbf{a}) Schematic of oblique plane light sheet microscope imaging a stained flagellum extracted from \emph{E. coli}. The light sheet rapidly scanned through the sample to image a 3D volume.
    (\textbf{b}) Representative center-of-mass displacement tracks along the lab axes for a diffusing flagellum.
    (\textbf{c}) Illustration of the three unit vectors ($\uvec{n}_1, \uvec{n}_2, \uvec{n}_3$) that describe the arbitrary helix orientation relative to the lab axes ($x,y,z$).
    (\textbf{d}) Orthographic projections of a 3D image of a single flagellum. Yellow circles and boxes indicate the same flagellum ends in three orthogonal views. Images are shown with a gamma of \num{0.75}.
\label{fig:figure-2}}
\end{figure*}

Over the past 120 years, significant effort has gone into improving methods for and applying FDT principles across various fields, including condensed matter, soft matter, and biological physics. As a result, the application of FDT is well understood and mature for spherical particles. In contrast, more complicated, asymmetric particle shapes introduce rich new physics, including non-Gaussian Brownian motion~\cite{Han2006} or propulsive coupling between rotational and translational motion~\cite{purcell1977life} (Fig.~\ref{fig:figure-1}b). To fully describe the position and orientation of an arbitrary-shaped particle, six total coordinates are required. The expanded parameter space allows for a more complicated set of ``fluctuations'' described as a matrix of correlation functions between displacements and rotations along the six coordinates. 

The corresponding dissipation function takes the form of a matrix of drag coefficients, often called the \emph{propulsion matrix}. The propulsion matrix describes fluid resistance to translations and rotations and fully characterizes the complex motility at low $\re$. The existence of the propulsion matrix is guaranteed at low $\re$ because the non-linear Navier-Stokes equations reduce to the linear Stokes equations~\cite{landau_fluids,Brennen1977}, relating applied forces ($\vec{F}$) and torques ($\tau$) and the particle's linear ($\vec{v}$) and angular ($\omega$) velocities,
\begin{equation}
\begin{pmatrix} 
    \vec{F} \\
    \vec{\tau}
\end{pmatrix} =
\begin{pmatrix}
    K_T & K_C^t \\
    K_C & K_R
\end{pmatrix}
\begin{pmatrix} 
    \vec{v} \\
    \vec{\omega}
\end{pmatrix}.
\label{eqn:1-propMatrix}
\end{equation}
Submatrices $K_T$ and $K_R$ represent translational and rotational drag, respectively, while $K_C$ describes the propulsive coupling between rotation and translation. As in the spherical case, the propulsion matrix coefficients depend only on the medium's viscosity, the particle's geometry, and the fluid boundary. The FDT connection between the propulsion matrix ($\vec{P}$) and the correlation function matrix ($\vec{D}$) described above was first given by Brenner, who showed it takes the form of a generalized Einstein relation, $\vec{D} = k_B T \vec{P}^{-1}$~\cite{Brenner1967}.

A key insight is that the propulsion matrix characterizes the efficiency of a rigid body at low $\re$ when acting as a propeller~\cite{purcell1997efficiency}. To obtain the propulsion matrix elements for any propeller geometry, it is possible to solve the Stokes equations using the Oseen tensor and flow singularities. However, the complexity of the boundary conditions makes analytic approaches intractable for all but the simplest geometries (e.g. spheres in a quasi-infinite fluid)~\cite{lauga2009hydrodynamics}. Similarly, experimental validation of theoretical predictions for complex geometries, such as helical particles, has not been possible due to the limited spatio-temporal bandwidth of available microscopy methods \cite{Kraft2013,bianchi2020brownian}. Instead, prior work quantified the propulsion matrix directly from the hydrodynamic definition using larger helical particles, scaling laws, and applied external forces (Fig.~\ref{fig:figure-1}c-e)~\cite{purcell1977life,chattopadhyay2006swimming,yuan2007measurement,rodenborn2013propulsion}.

Understanding the propulsion properties of asymmetric particles in low $\re$, where viscous forces dominate over inertial forces, is important because bacteria and other microscale swimmers live in this regime. At low $\re$, reciprocal motion cannot generate the net displacement required for propulsion~\cite{purcell1977life,lauga2009hydrodynamics}. As such, microscale swimmers must adopt different swimming strategies from larger organisms. For example, some swimmers have evolved to have flagella, propellers made out of filaments that generate translation by rotation~\cite{Berg1973}. Other swimmers utilize cilia, propellers made of filaments with a rigid power stroke to generate translation and highly flexible, disordered return stroke to avoid reciprocal motion~\cite{Katoh2018}. Understanding the efficiency of these molecular propellers is an outstanding question in biology and fluid mechanics, dating back over half a century~\cite{gray1955propulsion,lighthill1976flagellar}.

Here, we present direct quantification of the propulsion matrix for one type of microscale helical propeller, specifically the \emph{E. coli} flagellum. We measure the Brownian fluctuations of individual flagella using oblique plane microscopy (OPM)~\cite{dunsby2008optically}. We pioneer the capability of quantifying rotational diffusion along the longitudinal axis at the microscale and demonstrate that our theoretical analysis and computational framework (Supplemental Section~\ref{SI:sec-theory}-\ref{SI:sec-data-analysis}) obtain propulsion matrix coefficients comparable to those measured by conventional hydrodynamic methods~\cite{purcell1977life,chattopadhyay2006swimming,yuan2007measurement,rodenborn2013propulsion}. Our experimental measurements directly quantify the codiffusion coefficient that describes the coupling between the translation and rotation of the helix, confirming previous theoretical predictions about helical particle diffusion.

To begin, we generated volumetric timelapse movies of freely diffusing, isolated, and Cy3B-labeled helical flagella separated from an \emph{E. coli} body using a high-resolution OPM with fast volume imaging capability~\cite{dunsby2008optically,sapoznik2020versatile} (Fig.~\ref{fig:figure-2}a). The unique combination of fast and remote imaging, optical sectioning, and high spatial resolution provided by OPM enabled us to resolve the helix rotation without imparting inertia to the sample. As a result, our measurements obtained sharp 3D images of flagella (Fig.~\ref{fig:figure-2}d, Mov. S1-S2), providing dynamic information over a few tens of seconds (Fig.~\ref{fig:figure-2}b).

\begin{figure*}[ht]
\centering
\includegraphics[width=1\linewidth]{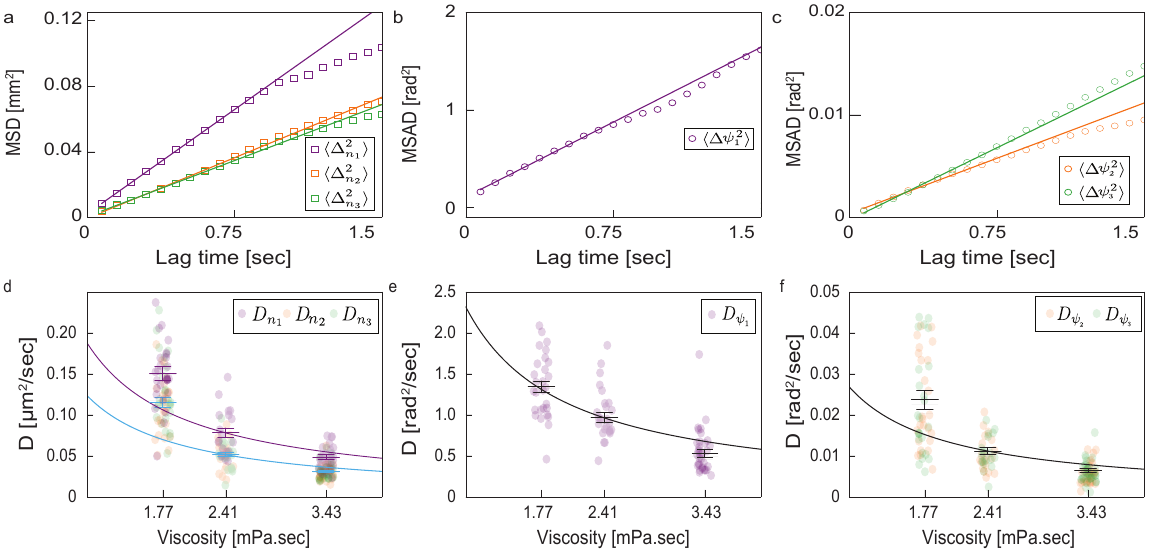}
\caption{
    \textbf{Diffusion coefficients along 6 degrees of freedom for flagella in different viscosities.}
    (\textbf{a}) Representative MSDs for displacement along the longitudinal (purple) and transverse (orange and green) axes. Solid lines are linear fits to the first ten data points.
    (\textbf{b}, \textbf{c}) Representative MSADs for rotation along the longitudinal (purple) and transverse (orange and green) axes.
    (\textbf{d}) Translational diffusion coefficients along the longitudinal (purple) and transverse (orange and green) axes in indicated viscosities. 
    (\textbf{e},\textbf{f}) Rotational diffusion coefficients about the transverse (purple) and longitudinal (orange and green) axes. Diffusion coefficients determined from each measurement (circles) and average diffusion coefficients (pluses) are shown. Error bars are the standard error of the mean (sem). Lines are fits to the expected $1/\eta$ viscosity scaling of the diffusion coefficients.
\label{fig:figure-3}}
\end{figure*}

To quantify the 3D motion of each helix, we parameterized the position using center-of-mass coordinates and orientation with three unit vectors or body axes, $\uvec{n}_1$, $\uvec{n}_2$, and $\uvec{n}_3$ (Fig.~\ref{fig:figure-2}c). Only three vector components are independent, which, when combined with the center-of-mass coordinates, implies that 6 degrees of freedom fully specify the helix position and orientation. Using each helix's position and orientation information, we calculated the diffusion matrix coefficients from the mean squared displacement (MSD), mean squared angular displacement (MSAD), or an appropriate generalization for each correlation function (Fig.~\ref{fig:figure-3}c,e, Fig.~\ref{fig:SI-fitting}-\ref{fig:SI-all-diffusion}, Alg.~\ref{alg:SI-MSD-trans}, Mov. S3, and Supplemental Section~\ref{SI:sec-data-analysis}-\ref{SI:sec-wall}).

Unlike spherical particles, asymmetric particles exhibit different diffusion coefficients depending on the relative direction of movement to the body axes. Due to the constant tumbling of the helix, the propulsion matrix expressions above apply only in the body frame. In the lab frame, tracks such as those plotted in Fig.~\ref{fig:figure-2}c will show different diffusive behavior depending on if the observation time is short or long compared with the tumbling time, $\tau_t$. For example, although a helix always diffuses anisotropically in the body frame, anisotropic diffusion can only be observed in the lab frame over times short compared with $\tau_t$, where the axis $\uvec{n}_1$ remains nearly fixed (Supplemental Section~\ref{SI:sec-simulation}). On longer time scales, the helix center of mass will diffuse with the rotationally-averaged diffusion coefficient~\cite{Han2006}.

For helical particles, symmetry considerations ensure that the propulsion matrix has only one off-diagonal term (Supplemental Section~\ref{SI:sec-symmetry}). Therefore, a flagellum is expected to undergo independent translational and rotational diffusion about its transverse axes, $\uvec{n}_2$ and $\uvec{n}_3$, and coupled translational and rotational diffusion about its longitudinal axis, $\uvec{n}_1$. Approximate screw symmetry implies that the diffusion coefficients for the transverse axes will be nearly equal. In this case, the propulsion matrix drag equations reduce to four uncoupled equations and one coupled equation of the form,
\begin{equation}
\begin{pmatrix} 
    F_1 \\
    \tau_1
\end{pmatrix} =
\begin{pmatrix}
    A & B \\
    B & D
\end{pmatrix}
\begin{pmatrix} 
    v_1 \\
    \omega_1
\end{pmatrix},
\label{eqn:1-propMatrix-purcell}
\end{equation}
where the forces, torques, and velocities all point along the helix's longitudinal axis.

Using the FDT, we obtain the propulsion matrix coefficients in terms of the diffusion coefficients $D_{n_1}$ and $D_{\psi_1}$ and the codiffusion coefficient $D_{n_1 \psi_1}$ (Supplemental Section \ref{SI:sec-theory})
\begin{flalign}
    A &= \frac{D_{\psi_1}}{D_{n_1} D_{\psi_1} - D_{n_1 \psi_1}^2} k_B T \label{eq:A}\\
    B &= - \frac{D_{n_1 \psi_1}}{D_{n_1} D_{\psi_1} - D_{n_1 \psi_1}^2} k_B T \label{eq:B}\\
    D &= \frac{D_{n_1}}{D_{n_1} D_{\psi_1} - D_{n_1 \psi_1}^2}  k_B T. \label{eq:D}
\end{flalign}
When propulsive coupling is absent, $D_{n_1 \psi_1} = 0$, eqs.~\ref{eq:A}--\ref{eq:D} recover the familiar Einstein relations $A = k_B T/D_{n_1}$ and $D = k_B T / D_{\psi_1}$.

We measured the diffusion coefficients' variation with viscosity using three concentrations of sucrose. Due to the linearity of the Stokes equations, we expect that all diffusion matrix coefficients scale inversely with viscosity $D \sim 1/ \eta$. Intuitively, a more viscous fluid impedes the flagella movement leading to reduced motion. As expected, the translational (Fig.~\ref{fig:figure-3}b) and rotational diffusion coefficients (Fig.~\ref{fig:figure-3}d,f) decrease with increasing viscosity. However, the diffusion coefficients for the lowest viscosity value were somewhat larger than expected. This is likely due to the uncertainty in determining $\eta$ when pipetting small volumes (tens of $\mu$L) of the highly viscous sucrose mixture (Fig. \ref{fig:SI-vis-diff}, Supplemental Section~\ref{SI:sec-viscosity}). As the diffusion coefficients may also depend on flagella length, we only considered flagella with lengths between \SIrange[range-phrase=~--~]{6}{10}{\micro \meter} and average length of \SI{7.77(91)}{\micro \meter} ($n=77$).

We found that flagella diffuse faster along the longitudinal axis than the transverse axes for all viscosities (Fig.~\ref{fig:figure-3}b). We quantified the anisotropy of the translational movement of the flagella by considering the ratio of the longitudinal to mean transverse diffusion coefficient, which were \num{1.33(07)}, \num{1.59(13)}, and \num{1.62(10)} for the three viscosity values considered. These results differ from the factor of two anticipated by resistive force theory (RFT), a qualitative approximation that replaces the fluid with two phenomenological drag coefficients~\cite{gray1955propulsion,lighthill1976flagellar,childress1981mechanics,rodenborn2013propulsion}.

We further found that the rotational diffusion along the longitudinal axis is two orders of magnitude faster than that along the transverse axis (Fig.~\ref{fig:figure-3}d,f). RFT again provides qualitative insight into the origin of this anisotropy. The fluid drag on a body is greater along the normal direction than in the transverse direction \cite{gray1955propulsion}. When the helix rotates along the transverse axes, it presents a larger normal surface area than when it rotates along the longitudinal axis and hence experiences greater resistance. The transverse rotational diffusion coefficients determine the tumbling time scales \cite{Han2006} $\tau_t = \SI{1}{\radian^2}/ 2 D_\perp \sim$ \num{20}, \num{40}, and \SI{100}{\second} respectively which are of similar order to the length of our full tracks.

\begin{figure*}[ht]
\centering
\includegraphics[width=1\linewidth]{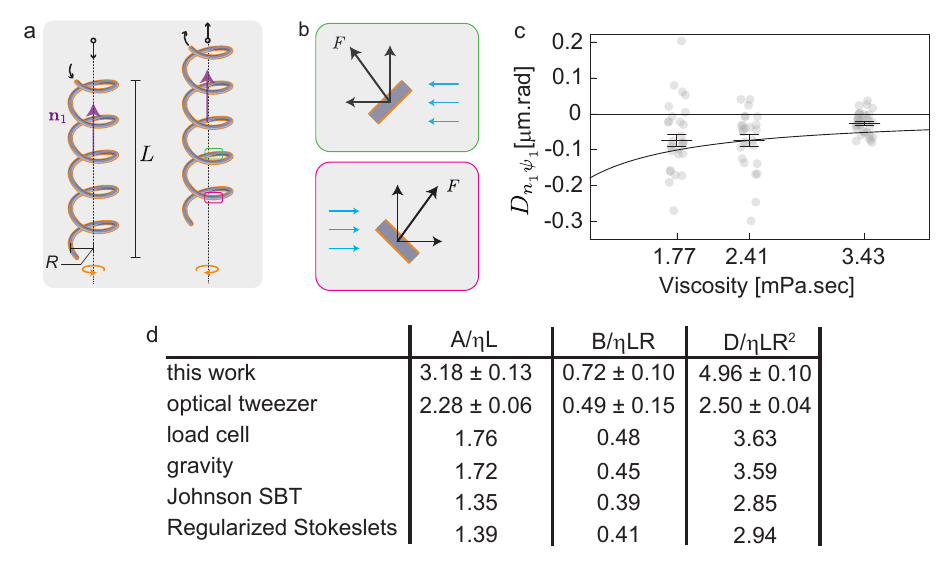}
\caption{
    \textbf{Correlation between translation and rotation.}  
    ({\bf a}) Schematic of translation induced by rotation of a left-handed helix in a viscous fluid. Counter-clockwise and clockwise rotations cause the helix to move downwards (left) and upwards (right), respectively.
    ({\bf b}) Free-body diagram of filament segments in their rest frames while helix rotates clockwise (panel a, right). Due to the rotation the segments see fluid flow (blue) in opposite directions. The forces in the horizontal direction cancel while those in the vertical direction add causing the helix to translate in the $\uvec{n}_1$ direction (upwards). Green and pink insets correspond to the regions shown in a.
    ({\bf c}) Codiffusion coefficients, $D_{n_1\psi_1}$, for flagella in different viscosities. Diffusion coefficients determined from each measurement (circles) and average diffusion coefficients (pluses) are shown. Error bars are sem.
    ({\bf d}) Non-dimensionalized propulsion matrix elements obtained from Brownian motion versus conventional hydrodynamics setup for helices with $\theta \sim \ang{32}$.
\label{fig:figure-4}}
\end{figure*}

The codiffusion coefficients describe the correlation between a flagellum's rotational and translational motion. In principle, there can be correlations between any pair of axes (Fig.~\ref{fig:SI-all-diffusion}, Supplemental Section~\ref{SI:sec-symmetry}). However, for a helix, we expect the dominant correlation between rotation about and translation along $\uvec{n}_1$. Due to its chiral shape, a helix tends to move in one direction when it rotates in a positive sense and the opposite direction when it rotates in a negative sense (Fig.~\ref{fig:figure-4}a,b). Unlike the translational and rotational diffusion coefficients, the codiffusion coefficient is not strictly positive, and the sign indicates the directional coupling between translational and rotational motion. Our measured codiffusion coefficients are negative, indicating that when the helix rotates in the right-handed sense about $\uvec{n}_1$ it translates along the $-\uvec{n}_1$ direction(Fig.~\ref{fig:figure-4}c). The negative value is expected as \emph{E. coli} flagella have a left-handed helical shape that rotates counterclockwise during steady motion when viewed from outside the cell~\cite{berg2003rotary, Kumar2009}. If the vector $\uvec{n}_1$ points toward the cell, the motor rotates the flagella ``negatively", driving the cell forward.

With these experimentally derived diffusion coefficients, we determined the coefficients of the propulsion matrix using Eqs.~\ref{eq:A}--\ref{eq:D}. The propulsion matrix coefficients are expected to depend only on the viscosity and helical geometry, and hence can be parameterized by the helical radius $R$, length $L$, filament radius $a$, and helical pitch $\lambda$. To allow comparison of our work with prior results, we define non-dimensionalized propulsion matrix coefficients $A^* \equiv A / \eta L$, $B^* \equiv B / \eta LR$, and $D^* \equiv D / \eta LR^2$ which remove all dependence on viscosity and first-order dependence on the flagellar length and helical diameter based on RFT. We note these non-dimensional coefficients do not remove dependence on the helical pitch (equivalently $R / \lambda$) or $\lambda /a$.

To determine the non-dimensional propulsion matrix elements from our measured diffusion coefficients, we relied on experimentally measured geometric parameters of \emph{E. coli} flagella. From image analysis of the $n = 88$ flagella, we find $l = \SI{7.7(14)}{\micro\meter}$ and $\lambda = \SI{2.5(1)}{\micro \meter}$, values that are consistent with previous measurements~\cite{turner2000real,Darnton2007}. Unlike in the viscosity experiments ($n=77$) we consider flagella of all lengths here. As the helical radius is on the order of the diffraction limit, we rely on the previously reported value $R = \SI{0.25}{\micro \meter}$, which implies that the helical pitch angle, $\theta = \tan^{-1}\left( 2\pi R / \lambda \right)$, is $\theta = \ang{32}$. As the filament radius is much smaller than the diffraction limit in our setup, we rely on the value obtained by x-ray diffraction, $a = \SI{0.01}{\micro \meter}$~\cite{Namba1989,Samatey2001}.

We compared non-dimensionalized propulsion matrix elements obtained from Brownian motion versus conventional hydrodynamics setup for helices with $\theta \sim \ang{32}$ (Fig.~\ref{fig:figure-4}d). We considered values from three experimental setups and two theory techniques~\cite{chattopadhyay2006swimming,rodenborn2013propulsion,purcell1997efficiency}. The mm-scale helix experiments agree and show reasonably close agreement with slender body theory (SBT) and regularized Stokeslet calculations~\cite{Johnson1980, Cortez2001, rodenborn2013propulsion}. Some of the remaining deviation between the theory and these experiments comes from using $\lambda / a \sim \num{200}$ which is accurate for the flagella but somewhat too small for the wire helices $\lambda / a \sim \num{40}$~\cite{rodenborn2013propulsion}.

The discrepancies between our results and the optical tweezer results may be related to an effective change in helix shape and mechanical properties of the flagellar bundle of \emph{E. coli}. The discrepancy is most significant in $D^*$, because $D^*$ as measured contains both \emph{E. coli} body flagellar drag coefficients. The two drag coefficients differ by an order of magnitude, leading to a potential systematic uncertainty in extracting only the flagellar drag coefficient. Additionally, this approach makes the RFT assumption that the total drag coefficient is the sum of the \emph{E. coli} body and the flagella, neglecting the hydrodynamic interactions between the body and the flagella~\cite{Tabak2014}.

Another interesting property that we can extract from the propulsion matrix is the maximum propeller efficiency, typically quantified as the ratio of the power that would be required to pull the object divided by the power required to propel it using a rotary motor. This efficiency is related to the propulsion matrix coefficients by $\varepsilon = B^{*2}/(4A^* D^*)$. $\varepsilon$ only depends on the shape of the propeller, independent of the cargo~\cite{chattopadhyay2006swimming,purcell1997efficiency} such as the \emph{E. coli} body and viscosity. Our measurement finds that the propulsion efficiency is $\varepsilon = \SI{1.7\pm0.22}{\percent}$ which is close to both the theoretical values of \SIrange{1}{3}{\percent}~\cite{lighthill1976flagellar,childress1981mechanics} and other experimental measurements~\cite{purcell1977life, chattopadhyay2006swimming}. 

It is initially surprising that the propulsion efficiency is so low. Perhaps, this is why \emph{E. coli} developed a highly efficient motor that can rotate at up to \num{18000} RPM~\cite{chen2000torque} with an \SI{\sim 80}{\percent} energy conversion rate during swimming~\cite{li2006low}. The motor allows \emph{E. coli} to travel at speeds of up to \SI[per-mode=symbol]{30}{\micro\metre\per\second}, about \num{15} body lengths per second, despite the low efficiency of the propeller. Some authors have argued that the low propulsion efficiency is unimportant because \emph{E. coli}'s power expenditure during swimming is estimated to be on the order of \SI[retain-zero-exponent]{e-16}{\watt}~\cite{chattopadhyay2006swimming,purcell1977life}, a small fraction of its total metabolic power consumption \SI[retain-zero-exponent]{e-13}{\watt}~\cite{Deng2021}. Recent work suggests that the metabolic cost of assembling a propeller may be equally critical. For example, the estimated raw material and energy cost to assemble an \emph{E. coli} flagellum are significantly less than Eukaryotic cilia~\cite{Schavemaker2022}.

In this work, we characterized the hydrodynamic properties of an isolated helical flagellum by measuring its propulsion matrix in the low Reynolds number regime. We combined recent advances in high-resolution, high-speed volumetric fluorescence imaging with a novel theoretical and data analysis approach that relies on the fluctuation-dissipation theorem and computational tracking of extended 3D particles. Our work introduces a new and general method for characterizing molecular propellers without the need to enforce external fluid flows or forces, as previously required. The approach we detail here has broad applications for studying the propulsion of other bacterial species and Eukaryotic flagella, as well as potential extensions to understanding hydrodynamic interactions between systems of multiple bacteria, multiple propellers, and the design of artificial microswimmers for targeted drug delivery and other medical applications~\cite{Nelson2010}.

\section*{Acknowledgments}
We thank A.A. Shrivastava and N.K. Ratheesh for providing MG1655WT plasmid and D. Gandavadi for preparing the microtubule samples. We thank G.B.M. Wisna for useful discussions and manuscript review.

RH acknowledges support from the National Institutes of Health Director’s New Innovator Award 1DP2AI144247, the National Science Foundation 2027215, and the Arizona Biomedical Research Consortium ADHS17-00007401.

DPS acknowledges support from Scialog, Research Corporation for Science Advancement, and Frederick Gardner Cottrell Foundation 28041 and the Chan Zuckerberg Initiative 2021-236170(5022)

\section*{Author Contributions}
\noindent{FD, PTB, BY, RFH, and DPS designed research; FD, PTB, DG, AC, JS, and DPS performed experiments; FD, PTB, BY, RFH, and DPS contributed new reagents/analytic tools; FD, PTB, BN, SR, BY, RFH, and DPS analyzed data; and FD, PTB, RFH, and DPS wrote the paper.}

\section*{Competing interests}
The authors declare they have no competing interests.

\section*{Data availability}
Microscope control and processing code is available at \url{https://github.com/QI2lab/OPM}. Data analysis code is available at \url{https://github.com/fdjutant/6-DOF-Flagella} and the version used in this manuscript is archived on Zenodo \url{https://doi.org/10.5281/zenodo.6562089}. All data is available from the corresponding author upon reasonable request.

\vfill


\bibliography{scibib}

\begin{thebibliography}{61}%
\makeatletter
\providecommand \@ifxundefined [1]{%
 \@ifx{#1\undefined}
}%
\providecommand \@ifnum [1]{%
 \ifnum #1\expandafter \@firstoftwo
 \else \expandafter \@secondoftwo
 \fi
}%
\providecommand \@ifx [1]{%
 \ifx #1\expandafter \@firstoftwo
 \else \expandafter \@secondoftwo
 \fi
}%
\providecommand \natexlab [1]{#1}%
\providecommand \enquote  [1]{``#1''}%
\providecommand \bibnamefont  [1]{#1}%
\providecommand \bibfnamefont [1]{#1}%
\providecommand \citenamefont [1]{#1}%
\providecommand \href@noop [0]{\@secondoftwo}%
\providecommand \href [0]{\begingroup \@sanitize@url \@href}%
\providecommand \@href[1]{\@@startlink{#1}\@@href}%
\providecommand \@@href[1]{\endgroup#1\@@endlink}%
\providecommand \@sanitize@url [0]{\catcode `\\12\catcode `\$12\catcode
  `\&12\catcode `\#12\catcode `\^12\catcode `\_12\catcode `\%12\relax}%
\providecommand \@@startlink[1]{}%
\providecommand \@@endlink[0]{}%
\providecommand \url  [0]{\begingroup\@sanitize@url \@url }%
\providecommand \@url [1]{\endgroup\@href {#1}{\urlprefix }}%
\providecommand \urlprefix  [0]{URL }%
\providecommand \Eprint [0]{\href }%
\providecommand \doibase [0]{https://doi.org/}%
\providecommand \selectlanguage [0]{\@gobble}%
\providecommand \bibinfo  [0]{\@secondoftwo}%
\providecommand \bibfield  [0]{\@secondoftwo}%
\providecommand \translation [1]{[#1]}%
\providecommand \BibitemOpen [0]{}%
\providecommand \bibitemStop [0]{}%
\providecommand \bibitemNoStop [0]{.\EOS\space}%
\providecommand \EOS [0]{\spacefactor3000\relax}%
\providecommand \BibitemShut  [1]{\csname bibitem#1\endcsname}%
\let\auto@bib@innerbib\@empty
\bibitem [{\citenamefont {Brown}(1828)}]{Brown1828}%
  \BibitemOpen
  \bibfield  {author} {\bibinfo {author} {\bibfnamefont {R.}~\bibnamefont
  {Brown}},\ }\bibfield  {title} {\bibinfo {title} {{XXVII}. a brief account of
  microscopical observations made in the months of {June}, {July} and {August}
  1827, on the particles contained in the pollen of plants; and on the general
  existence of active molecules in organic and inorganic bodies},\ }\href
  {https://doi.org/10.1080/14786442808674769} {\bibfield  {journal} {\bibinfo
  {journal} {The Philosophical Magazine}\ }\textbf {\bibinfo {volume} {4}},\
  \bibinfo {pages} {161} (\bibinfo {year} {1828})}\BibitemShut {NoStop}%
\bibitem [{\citenamefont {Perrin}(1909)}]{Perrin1909}%
  \BibitemOpen
  \bibfield  {author} {\bibinfo {author} {\bibfnamefont {J.}~\bibnamefont
  {Perrin}},\ }\bibfield  {title} {\bibinfo {title} {Mouvement brownien et
  réalité moléculaire},\ }\href@noop {} {\bibfield  {journal} {\bibinfo
  {journal} {Annales de chime et de physique}\ }\textbf {\bibinfo {volume}
  {18}},\ \bibinfo {pages} {1} (\bibinfo {year} {1909})}\BibitemShut {NoStop}%
\bibitem [{\citenamefont {Einstein}(1905)}]{einstein1905erzeugung}%
  \BibitemOpen
  \bibfield  {author} {\bibinfo {author} {\bibfnamefont {A.}~\bibnamefont
  {Einstein}},\ }\bibfield  {title} {\bibinfo {title} {{\"U}ber einem die
  erzeugung und verwandlung des lichtes betreffenden heuristischen
  gesichtspunkt},\ }\href@noop {} {\bibfield  {journal} {\bibinfo  {journal}
  {Annalen der physik}\ }\textbf {\bibinfo {volume} {4}} (\bibinfo {year}
  {1905})}\BibitemShut {NoStop}%
\bibitem [{\citenamefont {Kubo}(1966)}]{kubo1966fluctuation}%
  \BibitemOpen
  \bibfield  {author} {\bibinfo {author} {\bibfnamefont {R.}~\bibnamefont
  {Kubo}},\ }\bibfield  {title} {\bibinfo {title} {The fluctuation-dissipation
  theorem},\ }\href@noop {} {\bibfield  {journal} {\bibinfo  {journal} {Reports
  on progress in physics}\ }\textbf {\bibinfo {volume} {29}},\ \bibinfo {pages}
  {255} (\bibinfo {year} {1966})}\BibitemShut {NoStop}%
\bibitem [{\citenamefont {Chattopadhyay}\ \emph {et~al.}(2006)\citenamefont
  {Chattopadhyay}, \citenamefont {Moldovan}, \citenamefont {Yeung},\ and\
  \citenamefont {Wu}}]{chattopadhyay2006swimming}%
  \BibitemOpen
  \bibfield  {author} {\bibinfo {author} {\bibfnamefont {S.}~\bibnamefont
  {Chattopadhyay}}, \bibinfo {author} {\bibfnamefont {R.}~\bibnamefont
  {Moldovan}}, \bibinfo {author} {\bibfnamefont {C.}~\bibnamefont {Yeung}},\
  and\ \bibinfo {author} {\bibfnamefont {X.}~\bibnamefont {Wu}},\ }\bibfield
  {title} {\bibinfo {title} {Swimming efficiency of bacterium \emph{Escherichia
  coli}},\ }\href@noop {} {\bibfield  {journal} {\bibinfo  {journal}
  {Proceedings of the National Academy of Sciences}\ }\textbf {\bibinfo
  {volume} {103}},\ \bibinfo {pages} {13712} (\bibinfo {year}
  {2006})}\BibitemShut {NoStop}%
\bibitem [{\citenamefont {Rodenborn}\ \emph {et~al.}(2013)\citenamefont
  {Rodenborn}, \citenamefont {Chen}, \citenamefont {Swinney}, \citenamefont
  {Liu},\ and\ \citenamefont {Zhang}}]{rodenborn2013propulsion}%
  \BibitemOpen
  \bibfield  {author} {\bibinfo {author} {\bibfnamefont {B.}~\bibnamefont
  {Rodenborn}}, \bibinfo {author} {\bibfnamefont {C.-H.}\ \bibnamefont {Chen}},
  \bibinfo {author} {\bibfnamefont {H.~L.}\ \bibnamefont {Swinney}}, \bibinfo
  {author} {\bibfnamefont {B.}~\bibnamefont {Liu}},\ and\ \bibinfo {author}
  {\bibfnamefont {H.}~\bibnamefont {Zhang}},\ }\bibfield  {title} {\bibinfo
  {title} {Propulsion of microorganisms by a helical flagellum},\ }\href@noop
  {} {\bibfield  {journal} {\bibinfo  {journal} {Proceedings of the National
  Academy of Sciences}\ }\textbf {\bibinfo {volume} {110}},\ \bibinfo {pages}
  {E338} (\bibinfo {year} {2013})}\BibitemShut {NoStop}%
\bibitem [{\citenamefont {Han}\ \emph {et~al.}(2006)\citenamefont {Han},
  \citenamefont {Alsayed}, \citenamefont {Nobili}, \citenamefont {Zhang},
  \citenamefont {Lubensky},\ and\ \citenamefont {Yodh}}]{Han2006}%
  \BibitemOpen
  \bibfield  {author} {\bibinfo {author} {\bibfnamefont {Y.}~\bibnamefont
  {Han}}, \bibinfo {author} {\bibfnamefont {A.~M.}\ \bibnamefont {Alsayed}},
  \bibinfo {author} {\bibfnamefont {M.}~\bibnamefont {Nobili}}, \bibinfo
  {author} {\bibfnamefont {J.}~\bibnamefont {Zhang}}, \bibinfo {author}
  {\bibfnamefont {T.~C.}\ \bibnamefont {Lubensky}},\ and\ \bibinfo {author}
  {\bibfnamefont {A.~G.}\ \bibnamefont {Yodh}},\ }\bibfield  {title} {\bibinfo
  {title} {Brownian motion of an ellipsoid},\ }\href
  {https://doi.org/10.1126/science.1130146} {\bibfield  {journal} {\bibinfo
  {journal} {Science}\ }\textbf {\bibinfo {volume} {314}},\ \bibinfo {pages}
  {626} (\bibinfo {year} {2006})}\BibitemShut {NoStop}%
\bibitem [{\citenamefont {Purcell}(1977)}]{purcell1977life}%
  \BibitemOpen
  \bibfield  {author} {\bibinfo {author} {\bibfnamefont {E.~M.}\ \bibnamefont
  {Purcell}},\ }\bibfield  {title} {\bibinfo {title} {Life at low {Reynolds}
  number},\ }\href@noop {} {\bibfield  {journal} {\bibinfo  {journal} {American
  journal of physics}\ }\textbf {\bibinfo {volume} {45}},\ \bibinfo {pages} {3}
  (\bibinfo {year} {1977})}\BibitemShut {NoStop}%
\bibitem [{\citenamefont {Landau}\ and\ \citenamefont
  {Lifshitz}(2013)}]{landau_fluids}%
  \BibitemOpen
  \bibfield  {author} {\bibinfo {author} {\bibfnamefont {L.~D.}\ \bibnamefont
  {Landau}}\ and\ \bibinfo {author} {\bibfnamefont {E.~M.}\ \bibnamefont
  {Lifshitz}},\ }\href@noop {} {\emph {\bibinfo {title} {Fluid Mechanics}}},\
  \bibinfo {edition} {2nd}\ ed.\ (\bibinfo  {publisher} {Pergamon Press},\
  \bibinfo {address} {Oxford},\ \bibinfo {year} {2013})\BibitemShut {NoStop}%
\bibitem [{\citenamefont {Brennen}\ and\ \citenamefont
  {Winet}(1977)}]{Brennen1977}%
  \BibitemOpen
  \bibfield  {author} {\bibinfo {author} {\bibfnamefont {C.}~\bibnamefont
  {Brennen}}\ and\ \bibinfo {author} {\bibfnamefont {H.}~\bibnamefont
  {Winet}},\ }\bibfield  {title} {\bibinfo {title} {Fluid mechanics of
  propulsion by cilia and flagella},\ }\href
  {https://doi.org/10.1146/annurev.fl.09.010177.002011} {\bibfield  {journal}
  {\bibinfo  {journal} {Annual Review of Fluid Mechanics}\ }\textbf {\bibinfo
  {volume} {9}},\ \bibinfo {pages} {339} (\bibinfo {year} {1977})}\BibitemShut
  {NoStop}%
\bibitem [{\citenamefont {Brenner}(1967)}]{Brenner1967}%
  \BibitemOpen
  \bibfield  {author} {\bibinfo {author} {\bibfnamefont {H.}~\bibnamefont
  {Brenner}},\ }\bibfield  {title} {\bibinfo {title} {Coupling between the
  translational and rotational brownian motions of rigid particles of arbitrary
  shape},\ }\href {https://doi.org/10.1016/0021-9797(67)90185-3} {\bibfield
  {journal} {\bibinfo  {journal} {Journal of Colloid and Interface Science}\
  }\textbf {\bibinfo {volume} {23}},\ \bibinfo {pages} {407} (\bibinfo {year}
  {1967})}\BibitemShut {NoStop}%
\bibitem [{\citenamefont {Purcell}(1997)}]{purcell1997efficiency}%
  \BibitemOpen
  \bibfield  {author} {\bibinfo {author} {\bibfnamefont {E.~M.}\ \bibnamefont
  {Purcell}},\ }\bibfield  {title} {\bibinfo {title} {The efficiency of
  propulsion by a rotating flagellum},\ }\href@noop {} {\bibfield  {journal}
  {\bibinfo  {journal} {Proceedings of the National Academy of Sciences}\
  }\textbf {\bibinfo {volume} {94}},\ \bibinfo {pages} {11307} (\bibinfo {year}
  {1997})}\BibitemShut {NoStop}%
\bibitem [{\citenamefont {Lauga}\ and\ \citenamefont
  {Powers}(2009)}]{lauga2009hydrodynamics}%
  \BibitemOpen
  \bibfield  {author} {\bibinfo {author} {\bibfnamefont {E.}~\bibnamefont
  {Lauga}}\ and\ \bibinfo {author} {\bibfnamefont {T.~R.}\ \bibnamefont
  {Powers}},\ }\bibfield  {title} {\bibinfo {title} {The hydrodynamics of
  swimming microorganisms},\ }\href@noop {} {\bibfield  {journal} {\bibinfo
  {journal} {Reports on Progress in Physics}\ }\textbf {\bibinfo {volume}
  {72}},\ \bibinfo {pages} {096601} (\bibinfo {year} {2009})}\BibitemShut
  {NoStop}%
\bibitem [{\citenamefont {Kraft}\ \emph {et~al.}(2013)\citenamefont {Kraft},
  \citenamefont {Wittkowski}, \citenamefont {ten Hagen}, \citenamefont
  {Edmond}, \citenamefont {Pine},\ and\ \citenamefont {Löwen}}]{Kraft2013}%
  \BibitemOpen
  \bibfield  {author} {\bibinfo {author} {\bibfnamefont {D.~J.}\ \bibnamefont
  {Kraft}}, \bibinfo {author} {\bibfnamefont {R.}~\bibnamefont {Wittkowski}},
  \bibinfo {author} {\bibfnamefont {B.}~\bibnamefont {ten Hagen}}, \bibinfo
  {author} {\bibfnamefont {K.~V.}\ \bibnamefont {Edmond}}, \bibinfo {author}
  {\bibfnamefont {D.~J.}\ \bibnamefont {Pine}},\ and\ \bibinfo {author}
  {\bibfnamefont {H.}~\bibnamefont {Löwen}},\ }\bibfield  {title} {\bibinfo
  {title} {Brownian motion and the hydrodynamic friction tensor for colloidal
  particles of complex shape},\ }\href
  {https://doi.org/10.1103/physreve.88.050301} {\bibfield  {journal} {\bibinfo
  {journal} {Physical Review E}\ }\textbf {\bibinfo {volume} {88}},\ \bibinfo
  {pages} {050301} (\bibinfo {year} {2013})}\BibitemShut {NoStop}%
\bibitem [{\citenamefont {Bianchi}\ \emph {et~al.}(2020)\citenamefont
  {Bianchi}, \citenamefont {Sosa}, \citenamefont {Vizsnyiczai},\ and\
  \citenamefont {Di~Leonardo}}]{bianchi2020brownian}%
  \BibitemOpen
  \bibfield  {author} {\bibinfo {author} {\bibfnamefont {S.}~\bibnamefont
  {Bianchi}}, \bibinfo {author} {\bibfnamefont {V.~C.}\ \bibnamefont {Sosa}},
  \bibinfo {author} {\bibfnamefont {G.}~\bibnamefont {Vizsnyiczai}},\ and\
  \bibinfo {author} {\bibfnamefont {R.}~\bibnamefont {Di~Leonardo}},\
  }\bibfield  {title} {\bibinfo {title} {Brownian fluctuations and
  hydrodynamics of a microhelix near a solid wall},\ }\href@noop {} {\bibfield
  {journal} {\bibinfo  {journal} {Scientific reports}\ }\textbf {\bibinfo
  {volume} {10}},\ \bibinfo {pages} {1} (\bibinfo {year} {2020})}\BibitemShut
  {NoStop}%
\bibitem [{\citenamefont {Yuan}\ \emph {et~al.}(2007)\citenamefont {Yuan},
  \citenamefont {Liu},\ and\ \citenamefont {Yang}}]{yuan2007measurement}%
  \BibitemOpen
  \bibfield  {author} {\bibinfo {author} {\bibfnamefont {L.}~\bibnamefont
  {Yuan}}, \bibinfo {author} {\bibfnamefont {Z.}~\bibnamefont {Liu}},\ and\
  \bibinfo {author} {\bibfnamefont {J.}~\bibnamefont {Yang}},\ }\bibfield
  {title} {\bibinfo {title} {Measurement approach of brownian motion force by
  an abrupt tapered fiber optic tweezers},\ }\href@noop {} {\bibfield
  {journal} {\bibinfo  {journal} {Applied Physics Letters}\ }\textbf {\bibinfo
  {volume} {91}},\ \bibinfo {pages} {054101} (\bibinfo {year}
  {2007})}\BibitemShut {NoStop}%
\bibitem [{\citenamefont {Berg}\ and\ \citenamefont
  {Anderson}(1973)}]{Berg1973}%
  \BibitemOpen
  \bibfield  {author} {\bibinfo {author} {\bibfnamefont {H.~C.}\ \bibnamefont
  {Berg}}\ and\ \bibinfo {author} {\bibfnamefont {R.~A.}\ \bibnamefont
  {Anderson}},\ }\bibfield  {title} {\bibinfo {title} {Bacteria swim by
  rotating their flagellar filaments},\ }\href
  {https://doi.org/10.1038/245380a0} {\bibfield  {journal} {\bibinfo  {journal}
  {Nature}\ }\textbf {\bibinfo {volume} {245}},\ \bibinfo {pages} {380}
  (\bibinfo {year} {1973})}\BibitemShut {NoStop}%
\bibitem [{\citenamefont {Katoh}\ \emph {et~al.}(2018)\citenamefont {Katoh},
  \citenamefont {Ikegami}, \citenamefont {Uchida}, \citenamefont {Iwase},
  \citenamefont {Nakane}, \citenamefont {Masaike}, \citenamefont {Setou},\ and\
  \citenamefont {Nishizaka}}]{Katoh2018}%
  \BibitemOpen
  \bibfield  {author} {\bibinfo {author} {\bibfnamefont {T.~A.}\ \bibnamefont
  {Katoh}}, \bibinfo {author} {\bibfnamefont {K.}~\bibnamefont {Ikegami}},
  \bibinfo {author} {\bibfnamefont {N.}~\bibnamefont {Uchida}}, \bibinfo
  {author} {\bibfnamefont {T.}~\bibnamefont {Iwase}}, \bibinfo {author}
  {\bibfnamefont {D.}~\bibnamefont {Nakane}}, \bibinfo {author} {\bibfnamefont
  {T.}~\bibnamefont {Masaike}}, \bibinfo {author} {\bibfnamefont
  {M.}~\bibnamefont {Setou}},\ and\ \bibinfo {author} {\bibfnamefont
  {T.}~\bibnamefont {Nishizaka}},\ }\bibfield  {title} {\bibinfo {title}
  {Three-dimensional tracking of microbeads attached to the tip of single
  isolated tracheal cilia beating under external load},\ }\bibfield  {journal}
  {\bibinfo  {journal} {Scientific Reports}\ }\textbf {\bibinfo {volume} {8}},\
  \href {https://doi.org/10.1038/s41598-018-33846-5}
  {10.1038/s41598-018-33846-5} (\bibinfo {year} {2018})\BibitemShut {NoStop}%
\bibitem [{\citenamefont {Gray}\ and\ \citenamefont
  {Hancock}(1955)}]{gray1955propulsion}%
  \BibitemOpen
  \bibfield  {author} {\bibinfo {author} {\bibfnamefont {J.}~\bibnamefont
  {Gray}}\ and\ \bibinfo {author} {\bibfnamefont {G.}~\bibnamefont {Hancock}},\
  }\bibfield  {title} {\bibinfo {title} {The propulsion of sea-urchin
  spermatozoa},\ }\href@noop {} {\bibfield  {journal} {\bibinfo  {journal}
  {Journal of Experimental Biology}\ }\textbf {\bibinfo {volume} {32}},\
  \bibinfo {pages} {802} (\bibinfo {year} {1955})}\BibitemShut {NoStop}%
\bibitem [{\citenamefont {Lighthill}(1976)}]{lighthill1976flagellar}%
  \BibitemOpen
  \bibfield  {author} {\bibinfo {author} {\bibfnamefont {J.}~\bibnamefont
  {Lighthill}},\ }\bibfield  {title} {\bibinfo {title} {Flagellar
  hydrodynamics},\ }\href@noop {} {\bibfield  {journal} {\bibinfo  {journal}
  {SIAM review}\ }\textbf {\bibinfo {volume} {18}},\ \bibinfo {pages} {161}
  (\bibinfo {year} {1976})}\BibitemShut {NoStop}%
\bibitem [{\citenamefont {Dunsby}(2008)}]{dunsby2008optically}%
  \BibitemOpen
  \bibfield  {author} {\bibinfo {author} {\bibfnamefont {C.}~\bibnamefont
  {Dunsby}},\ }\bibfield  {title} {\bibinfo {title} {Optically sectioned
  imaging by oblique plane microscopy},\ }\href@noop {} {\bibfield  {journal}
  {\bibinfo  {journal} {Optics express}\ }\textbf {\bibinfo {volume} {16}},\
  \bibinfo {pages} {20306} (\bibinfo {year} {2008})}\BibitemShut {NoStop}%
\bibitem [{\citenamefont {Sapoznik}\ \emph {et~al.}(2020)\citenamefont
  {Sapoznik}, \citenamefont {Chang}, \citenamefont {Huh}, \citenamefont {Ju},
  \citenamefont {Azarova}, \citenamefont {Pohlkamp}, \citenamefont {Welf},
  \citenamefont {Broadbent}, \citenamefont {Carisey}, \citenamefont {Stehbens}
  \emph {et~al.}}]{sapoznik2020versatile}%
  \BibitemOpen
  \bibfield  {author} {\bibinfo {author} {\bibfnamefont {E.}~\bibnamefont
  {Sapoznik}}, \bibinfo {author} {\bibfnamefont {B.-J.}\ \bibnamefont {Chang}},
  \bibinfo {author} {\bibfnamefont {J.}~\bibnamefont {Huh}}, \bibinfo {author}
  {\bibfnamefont {R.~J.}\ \bibnamefont {Ju}}, \bibinfo {author} {\bibfnamefont
  {E.~V.}\ \bibnamefont {Azarova}}, \bibinfo {author} {\bibfnamefont
  {T.}~\bibnamefont {Pohlkamp}}, \bibinfo {author} {\bibfnamefont {E.~S.}\
  \bibnamefont {Welf}}, \bibinfo {author} {\bibfnamefont {D.}~\bibnamefont
  {Broadbent}}, \bibinfo {author} {\bibfnamefont {A.~F.}\ \bibnamefont
  {Carisey}}, \bibinfo {author} {\bibfnamefont {S.~J.}\ \bibnamefont
  {Stehbens}}, \emph {et~al.},\ }\bibfield  {title} {\bibinfo {title} {A
  versatile oblique plane microscope for large-scale and high-resolution
  imaging of subcellular dynamics},\ }\href@noop {} {\bibfield  {journal}
  {\bibinfo  {journal} {eLife}\ }\textbf {\bibinfo {volume} {9}},\ \bibinfo
  {pages} {e57681} (\bibinfo {year} {2020})}\BibitemShut {NoStop}%
\bibitem [{\citenamefont {Childress}(1981)}]{childress1981mechanics}%
  \BibitemOpen
  \bibfield  {author} {\bibinfo {author} {\bibfnamefont {S.}~\bibnamefont
  {Childress}},\ }\href {https://doi.org/10.1017/CBO9780511569593} {\emph
  {\bibinfo {title} {Mechanics of Swimming and Flying}}},\ Cambridge Studies in
  Mathematical Biology\ (\bibinfo  {publisher} {Cambridge University Press},\
  \bibinfo {address} {Cambridge},\ \bibinfo {year} {1981})\BibitemShut
  {NoStop}%
\bibitem [{\citenamefont {Berg}(2003)}]{berg2003rotary}%
  \BibitemOpen
  \bibfield  {author} {\bibinfo {author} {\bibfnamefont {H.~C.}\ \bibnamefont
  {Berg}},\ }\bibfield  {title} {\bibinfo {title} {The rotary motor of
  bacterial flagella},\ }\href@noop {} {\bibfield  {journal} {\bibinfo
  {journal} {Annual review of biochemistry}\ }\textbf {\bibinfo {volume}
  {72}},\ \bibinfo {pages} {19} (\bibinfo {year} {2003})}\BibitemShut {NoStop}%
\bibitem [{\citenamefont {Kumar}\ and\ \citenamefont
  {Philominathan}(2009)}]{Kumar2009}%
  \BibitemOpen
  \bibfield  {author} {\bibinfo {author} {\bibfnamefont {M.~S.}\ \bibnamefont
  {Kumar}}\ and\ \bibinfo {author} {\bibfnamefont {P.}~\bibnamefont
  {Philominathan}},\ }\bibfield  {title} {\bibinfo {title} {The physics of
  flagellar motion of \emph{E. coli} during chemotaxis},\ }\href
  {https://doi.org/10.1007/s12551-009-0024-5} {\bibfield  {journal} {\bibinfo
  {journal} {Biophysical Reviews}\ }\textbf {\bibinfo {volume} {2}},\ \bibinfo
  {pages} {13} (\bibinfo {year} {2009})}\BibitemShut {NoStop}%
\bibitem [{\citenamefont {Turner}\ \emph {et~al.}(2000)\citenamefont {Turner},
  \citenamefont {Ryu},\ and\ \citenamefont {Berg}}]{turner2000real}%
  \BibitemOpen
  \bibfield  {author} {\bibinfo {author} {\bibfnamefont {L.}~\bibnamefont
  {Turner}}, \bibinfo {author} {\bibfnamefont {W.~S.}\ \bibnamefont {Ryu}},\
  and\ \bibinfo {author} {\bibfnamefont {H.~C.}\ \bibnamefont {Berg}},\
  }\bibfield  {title} {\bibinfo {title} {Real-time imaging of fluorescent
  flagellar filaments},\ }\href@noop {} {\bibfield  {journal} {\bibinfo
  {journal} {Journal of bacteriology}\ }\textbf {\bibinfo {volume} {182}},\
  \bibinfo {pages} {2793} (\bibinfo {year} {2000})}\BibitemShut {NoStop}%
\bibitem [{\citenamefont {Darnton}\ \emph {et~al.}(2007)\citenamefont
  {Darnton}, \citenamefont {Turner}, \citenamefont {Rojevsky},\ and\
  \citenamefont {Berg}}]{Darnton2007}%
  \BibitemOpen
  \bibfield  {author} {\bibinfo {author} {\bibfnamefont {N.~C.}\ \bibnamefont
  {Darnton}}, \bibinfo {author} {\bibfnamefont {L.}~\bibnamefont {Turner}},
  \bibinfo {author} {\bibfnamefont {S.}~\bibnamefont {Rojevsky}},\ and\
  \bibinfo {author} {\bibfnamefont {H.~C.}\ \bibnamefont {Berg}},\ }\bibfield
  {title} {\bibinfo {title} {On torque and tumbling in swimming
  \emph{Escherichia coli}},\ }\href {https://doi.org/10.1128/jb.01501-06}
  {\bibfield  {journal} {\bibinfo  {journal} {Journal of Bacteriology}\
  }\textbf {\bibinfo {volume} {189}},\ \bibinfo {pages} {1756} (\bibinfo {year}
  {2007})}\BibitemShut {NoStop}%
\bibitem [{\citenamefont {Namba}\ \emph {et~al.}(1989)\citenamefont {Namba},
  \citenamefont {Yamashita},\ and\ \citenamefont {Vonderviszt}}]{Namba1989}%
  \BibitemOpen
  \bibfield  {author} {\bibinfo {author} {\bibfnamefont {K.}~\bibnamefont
  {Namba}}, \bibinfo {author} {\bibfnamefont {I.}~\bibnamefont {Yamashita}},\
  and\ \bibinfo {author} {\bibfnamefont {F.}~\bibnamefont {Vonderviszt}},\
  }\bibfield  {title} {\bibinfo {title} {Structure of the core and central
  channel of bacterial flagella},\ }\href {https://doi.org/10.1038/342648a0}
  {\bibfield  {journal} {\bibinfo  {journal} {Nature}\ }\textbf {\bibinfo
  {volume} {342}},\ \bibinfo {pages} {648} (\bibinfo {year}
  {1989})}\BibitemShut {NoStop}%
\bibitem [{\citenamefont {Samatey}\ \emph {et~al.}(2001)\citenamefont
  {Samatey}, \citenamefont {Imada}, \citenamefont {Nagashima}, \citenamefont
  {Vonderviszt}, \citenamefont {Kumasaka}, \citenamefont {Yamamoto},\ and\
  \citenamefont {Namba}}]{Samatey2001}%
  \BibitemOpen
  \bibfield  {author} {\bibinfo {author} {\bibfnamefont {F.~A.}\ \bibnamefont
  {Samatey}}, \bibinfo {author} {\bibfnamefont {K.}~\bibnamefont {Imada}},
  \bibinfo {author} {\bibfnamefont {S.}~\bibnamefont {Nagashima}}, \bibinfo
  {author} {\bibfnamefont {F.}~\bibnamefont {Vonderviszt}}, \bibinfo {author}
  {\bibfnamefont {T.}~\bibnamefont {Kumasaka}}, \bibinfo {author}
  {\bibfnamefont {M.}~\bibnamefont {Yamamoto}},\ and\ \bibinfo {author}
  {\bibfnamefont {K.}~\bibnamefont {Namba}},\ }\bibfield  {title} {\bibinfo
  {title} {Structure of the bacterial flagellar protofilament and implications
  for a switch for supercoiling},\ }\href {https://doi.org/10.1038/35066504}
  {\bibfield  {journal} {\bibinfo  {journal} {Nature}\ }\textbf {\bibinfo
  {volume} {410}},\ \bibinfo {pages} {331} (\bibinfo {year}
  {2001})}\BibitemShut {NoStop}%
\bibitem [{\citenamefont {Johnson}(1980)}]{Johnson1980}%
  \BibitemOpen
  \bibfield  {author} {\bibinfo {author} {\bibfnamefont {R.~E.}\ \bibnamefont
  {Johnson}},\ }\bibfield  {title} {\bibinfo {title} {An improved slender-body
  theory for {Stokes} flow},\ }\href
  {https://doi.org/10.1017/s0022112080000687} {\bibfield  {journal} {\bibinfo
  {journal} {Journal of Fluid Mechanics}\ }\textbf {\bibinfo {volume} {99}},\
  \bibinfo {pages} {411} (\bibinfo {year} {1980})}\BibitemShut {NoStop}%
\bibitem [{\citenamefont {Cortez}(2001)}]{Cortez2001}%
  \BibitemOpen
  \bibfield  {author} {\bibinfo {author} {\bibfnamefont {R.}~\bibnamefont
  {Cortez}},\ }\bibfield  {title} {\bibinfo {title} {The method of regularized
  stokeslets},\ }\href {https://doi.org/10.1137/s106482750038146x} {\bibfield
  {journal} {\bibinfo  {journal} {{SIAM} Journal on Scientific Computing}\
  }\textbf {\bibinfo {volume} {23}},\ \bibinfo {pages} {1204} (\bibinfo {year}
  {2001})}\BibitemShut {NoStop}%
\bibitem [{\citenamefont {Tabak}\ and\ \citenamefont
  {Yesilyurt}(2014)}]{Tabak2014}%
  \BibitemOpen
  \bibfield  {author} {\bibinfo {author} {\bibfnamefont {A.}~\bibnamefont
  {Tabak}}\ and\ \bibinfo {author} {\bibfnamefont {S.}~\bibnamefont
  {Yesilyurt}},\ }\bibfield  {title} {\bibinfo {title}
  {Computationally-validated surrogate models for optimal geometric design of
  bio-inspired swimming robots: Helical swimmers},\ }\href
  {https://doi.org/10.1016/j.compfluid.2014.04.033} {\bibfield  {journal}
  {\bibinfo  {journal} {Computers Fluids}\ }\textbf {\bibinfo {volume} {99}},\
  \bibinfo {pages} {190} (\bibinfo {year} {2014})}\BibitemShut {NoStop}%
\bibitem [{\citenamefont {Chen}\ and\ \citenamefont
  {Berg}(2000)}]{chen2000torque}%
  \BibitemOpen
  \bibfield  {author} {\bibinfo {author} {\bibfnamefont {X.}~\bibnamefont
  {Chen}}\ and\ \bibinfo {author} {\bibfnamefont {H.~C.}\ \bibnamefont
  {Berg}},\ }\bibfield  {title} {\bibinfo {title} {Torque-speed relationship of
  the flagellar rotary motor of \emph{Escherichia coli}},\ }\href@noop {}
  {\bibfield  {journal} {\bibinfo  {journal} {Biophysical journal}\ }\textbf
  {\bibinfo {volume} {78}},\ \bibinfo {pages} {1036} (\bibinfo {year}
  {2000})}\BibitemShut {NoStop}%
\bibitem [{\citenamefont {Li}\ and\ \citenamefont {Tang}(2006)}]{li2006low}%
  \BibitemOpen
  \bibfield  {author} {\bibinfo {author} {\bibfnamefont {G.}~\bibnamefont
  {Li}}\ and\ \bibinfo {author} {\bibfnamefont {J.~X.}\ \bibnamefont {Tang}},\
  }\bibfield  {title} {\bibinfo {title} {Low flagellar motor torque and high
  swimming efficiency of \emph{Caulobacter crescentus} swarmer cells},\
  }\href@noop {} {\bibfield  {journal} {\bibinfo  {journal} {Biophysical
  journal}\ }\textbf {\bibinfo {volume} {91}},\ \bibinfo {pages} {2726}
  (\bibinfo {year} {2006})}\BibitemShut {NoStop}%
\bibitem [{\citenamefont {Deng}\ \emph {et~al.}(2021)\citenamefont {Deng},
  \citenamefont {Beahm}, \citenamefont {Ionov},\ and\ \citenamefont
  {Sarpeshkar}}]{Deng2021}%
  \BibitemOpen
  \bibfield  {author} {\bibinfo {author} {\bibfnamefont {Y.}~\bibnamefont
  {Deng}}, \bibinfo {author} {\bibfnamefont {D.~R.}\ \bibnamefont {Beahm}},
  \bibinfo {author} {\bibfnamefont {S.}~\bibnamefont {Ionov}},\ and\ \bibinfo
  {author} {\bibfnamefont {R.}~\bibnamefont {Sarpeshkar}},\ }\bibfield  {title}
  {\bibinfo {title} {Measuring and modeling energy and power consumption in
  living microbial cells with a synthetic {ATP} reporter},\ }\bibfield
  {journal} {\bibinfo  {journal} {{BMC} Biology}\ }\textbf {\bibinfo {volume}
  {19}},\ \href {https://doi.org/10.1186/s12915-021-01023-2}
  {10.1186/s12915-021-01023-2} (\bibinfo {year} {2021})\BibitemShut {NoStop}%
\bibitem [{\citenamefont {Schavemaker}\ and\ \citenamefont
  {Lynch}(2022)}]{Schavemaker2022}%
  \BibitemOpen
  \bibfield  {author} {\bibinfo {author} {\bibfnamefont {P.~E.}\ \bibnamefont
  {Schavemaker}}\ and\ \bibinfo {author} {\bibfnamefont {M.}~\bibnamefont
  {Lynch}},\ }\bibfield  {title} {\bibinfo {title} {Flagellar energy costs
  across the tree of life},\ }\href {https://doi.org/10.7554/eLife.77266}
  {\bibfield  {journal} {\bibinfo  {journal} {eLife}\ }\textbf {\bibinfo
  {volume} {11}},\ \bibinfo {pages} {e77266} (\bibinfo {year}
  {2022})}\BibitemShut {NoStop}%
\bibitem [{\citenamefont {Nelson}\ \emph {et~al.}(2010)\citenamefont {Nelson},
  \citenamefont {Kaliakatsos},\ and\ \citenamefont {Abbott}}]{Nelson2010}%
  \BibitemOpen
  \bibfield  {author} {\bibinfo {author} {\bibfnamefont {B.~J.}\ \bibnamefont
  {Nelson}}, \bibinfo {author} {\bibfnamefont {I.~K.}\ \bibnamefont
  {Kaliakatsos}},\ and\ \bibinfo {author} {\bibfnamefont {J.~J.}\ \bibnamefont
  {Abbott}},\ }\bibfield  {title} {\bibinfo {title} {Microrobots for minimally
  invasive medicine},\ }\href
  {https://doi.org/10.1146/annurev-bioeng-010510-103409} {\bibfield  {journal}
  {\bibinfo  {journal} {Annual Review of Biomedical Engineering}\ }\textbf
  {\bibinfo {volume} {12}},\ \bibinfo {pages} {55} (\bibinfo {year}
  {2010})}\BibitemShut {NoStop}%
\bibitem [{\citenamefont {Hyman}\ \emph {et~al.}(1992)\citenamefont {Hyman},
  \citenamefont {Salser}, \citenamefont {Drechsel}, \citenamefont {Unwin},\
  and\ \citenamefont {Mitchison}}]{hyman1992role}%
  \BibitemOpen
  \bibfield  {author} {\bibinfo {author} {\bibfnamefont {A.~A.}\ \bibnamefont
  {Hyman}}, \bibinfo {author} {\bibfnamefont {S.}~\bibnamefont {Salser}},
  \bibinfo {author} {\bibfnamefont {D.}~\bibnamefont {Drechsel}}, \bibinfo
  {author} {\bibfnamefont {N.}~\bibnamefont {Unwin}},\ and\ \bibinfo {author}
  {\bibfnamefont {T.~J.}\ \bibnamefont {Mitchison}},\ }\bibfield  {title}
  {\bibinfo {title} {Role of {GTP} hydrolysis in microtubule dynamics:
  information from a slowly hydrolyzable analogue, {GMPCPP}.},\ }\href@noop {}
  {\bibfield  {journal} {\bibinfo  {journal} {Molecular biology of the cell}\
  }\textbf {\bibinfo {volume} {3}},\ \bibinfo {pages} {1155} (\bibinfo {year}
  {1992})}\BibitemShut {NoStop}%
\bibitem [{\citenamefont {Gell}\ \emph {et~al.}(2010)\citenamefont {Gell},
  \citenamefont {Bormuth}, \citenamefont {Brouhard}, \citenamefont {Cohen},
  \citenamefont {Diez}, \citenamefont {Friel}, \citenamefont {Helenius},
  \citenamefont {Nitzsche}, \citenamefont {Petzold}, \citenamefont {Ribbe}
  \emph {et~al.}}]{gell2010microtubule}%
  \BibitemOpen
  \bibfield  {author} {\bibinfo {author} {\bibfnamefont {C.}~\bibnamefont
  {Gell}}, \bibinfo {author} {\bibfnamefont {V.}~\bibnamefont {Bormuth}},
  \bibinfo {author} {\bibfnamefont {G.~J.}\ \bibnamefont {Brouhard}}, \bibinfo
  {author} {\bibfnamefont {D.~N.}\ \bibnamefont {Cohen}}, \bibinfo {author}
  {\bibfnamefont {S.}~\bibnamefont {Diez}}, \bibinfo {author} {\bibfnamefont
  {C.~T.}\ \bibnamefont {Friel}}, \bibinfo {author} {\bibfnamefont
  {J.}~\bibnamefont {Helenius}}, \bibinfo {author} {\bibfnamefont
  {B.}~\bibnamefont {Nitzsche}}, \bibinfo {author} {\bibfnamefont
  {H.}~\bibnamefont {Petzold}}, \bibinfo {author} {\bibfnamefont
  {J.}~\bibnamefont {Ribbe}}, \emph {et~al.},\ }\bibfield  {title} {\bibinfo
  {title} {Microtubule dynamics reconstituted in vitro and imaged by
  single-molecule fluorescence microscopy},\ }\href@noop {} {\bibfield
  {journal} {\bibinfo  {journal} {Methods in cell biology}\ }\textbf {\bibinfo
  {volume} {95}},\ \bibinfo {pages} {221} (\bibinfo {year} {2010})}\BibitemShut
  {NoStop}%
\bibitem [{\citenamefont {Bouchard}\ \emph {et~al.}(2015)\citenamefont
  {Bouchard}, \citenamefont {Voleti}, \citenamefont {Mendes}, \citenamefont
  {Lacefield}, \citenamefont {Grueber}, \citenamefont {Mann}, \citenamefont
  {Bruno},\ and\ \citenamefont {Hillman}}]{bouchard2015swept}%
  \BibitemOpen
  \bibfield  {author} {\bibinfo {author} {\bibfnamefont {M.~B.}\ \bibnamefont
  {Bouchard}}, \bibinfo {author} {\bibfnamefont {V.}~\bibnamefont {Voleti}},
  \bibinfo {author} {\bibfnamefont {C.~S.}\ \bibnamefont {Mendes}}, \bibinfo
  {author} {\bibfnamefont {C.}~\bibnamefont {Lacefield}}, \bibinfo {author}
  {\bibfnamefont {W.~B.}\ \bibnamefont {Grueber}}, \bibinfo {author}
  {\bibfnamefont {R.~S.}\ \bibnamefont {Mann}}, \bibinfo {author}
  {\bibfnamefont {R.~M.}\ \bibnamefont {Bruno}},\ and\ \bibinfo {author}
  {\bibfnamefont {E.~M.}\ \bibnamefont {Hillman}},\ }\bibfield  {title}
  {\bibinfo {title} {Swept confocally-aligned planar excitation ({SCAPE})
  microscopy for high-speed volumetric imaging of behaving organisms},\
  }\href@noop {} {\bibfield  {journal} {\bibinfo  {journal} {Nature photonics}\
  }\textbf {\bibinfo {volume} {9}},\ \bibinfo {pages} {113} (\bibinfo {year}
  {2015})}\BibitemShut {NoStop}%
\bibitem [{\citenamefont {Kumar}\ \emph {et~al.}(2018)\citenamefont {Kumar},
  \citenamefont {Kishore}, \citenamefont {Nasenbeny}, \citenamefont {McLean},\
  and\ \citenamefont {Kozorovitskiy}}]{kumar2018integrated}%
  \BibitemOpen
  \bibfield  {author} {\bibinfo {author} {\bibfnamefont {M.}~\bibnamefont
  {Kumar}}, \bibinfo {author} {\bibfnamefont {S.}~\bibnamefont {Kishore}},
  \bibinfo {author} {\bibfnamefont {J.}~\bibnamefont {Nasenbeny}}, \bibinfo
  {author} {\bibfnamefont {D.~L.}\ \bibnamefont {McLean}},\ and\ \bibinfo
  {author} {\bibfnamefont {Y.}~\bibnamefont {Kozorovitskiy}},\ }\bibfield
  {title} {\bibinfo {title} {Integrated one-and two-photon scanned oblique
  plane illumination ({SOPi}) microscopy for rapid volumetric imaging},\
  }\href@noop {} {\bibfield  {journal} {\bibinfo  {journal} {Optics Express}\
  }\textbf {\bibinfo {volume} {26}},\ \bibinfo {pages} {13027} (\bibinfo {year}
  {2018})}\BibitemShut {NoStop}%
\bibitem [{\citenamefont {Millett-Sikking}\ and\ \citenamefont
  {York}(2019)}]{york2019}%
  \BibitemOpen
  \bibfield  {author} {\bibinfo {author} {\bibfnamefont {A.}~\bibnamefont
  {Millett-Sikking}}\ and\ \bibinfo {author} {\bibfnamefont {A.}~\bibnamefont
  {York}},\ }\bibfield  {title} {\bibinfo {title} {High {NA} single-objective
  light-sheet},\ }\bibfield  {journal} {\bibinfo  {journal}
  {10.5281/zenodo.3376243}\ }\href {https://doi.org/10.5281/zenodo.3376243}
  {10.5281/zenodo.3376243} (\bibinfo {year} {2019})\BibitemShut {NoStop}%
\bibitem [{\citenamefont {Edelstein}\ \emph {et~al.}(2014)\citenamefont
  {Edelstein}, \citenamefont {Tsuchida}, \citenamefont {Amodaj}, \citenamefont
  {Pinkard}, \citenamefont {Vale},\ and\ \citenamefont
  {Stuurman}}]{edelstein2014advanced}%
  \BibitemOpen
  \bibfield  {author} {\bibinfo {author} {\bibfnamefont {A.~D.}\ \bibnamefont
  {Edelstein}}, \bibinfo {author} {\bibfnamefont {M.~A.}\ \bibnamefont
  {Tsuchida}}, \bibinfo {author} {\bibfnamefont {N.}~\bibnamefont {Amodaj}},
  \bibinfo {author} {\bibfnamefont {H.}~\bibnamefont {Pinkard}}, \bibinfo
  {author} {\bibfnamefont {R.~D.}\ \bibnamefont {Vale}},\ and\ \bibinfo
  {author} {\bibfnamefont {N.}~\bibnamefont {Stuurman}},\ }\bibfield  {title}
  {\bibinfo {title} {Advanced methods of microscope control using
  $\mu${M}anager software},\ }\href@noop {} {\bibfield  {journal} {\bibinfo
  {journal} {Journal of biological methods}\ }\textbf {\bibinfo {volume} {1}}
  (\bibinfo {year} {2014})}\BibitemShut {NoStop}%
\bibitem [{\citenamefont {Sofroniew}\ \emph {et~al.}(2022)\citenamefont
  {Sofroniew}, \citenamefont {Lambert}, \citenamefont {Evans}, \citenamefont
  {Nunez-Iglesias}, \citenamefont {Bokota}, \citenamefont {Winston},
  \citenamefont {Peña-Castellanos}, \citenamefont {Yamauchi}, \citenamefont
  {Bussonnier}, \citenamefont {Doncila~Pop},\ and\ \citenamefont
  {et~al.}}]{napari2022}%
  \BibitemOpen
  \bibfield  {author} {\bibinfo {author} {\bibfnamefont {N.}~\bibnamefont
  {Sofroniew}}, \bibinfo {author} {\bibfnamefont {T.}~\bibnamefont {Lambert}},
  \bibinfo {author} {\bibfnamefont {K.}~\bibnamefont {Evans}}, \bibinfo
  {author} {\bibfnamefont {J.}~\bibnamefont {Nunez-Iglesias}}, \bibinfo
  {author} {\bibfnamefont {G.}~\bibnamefont {Bokota}}, \bibinfo {author}
  {\bibfnamefont {P.}~\bibnamefont {Winston}}, \bibinfo {author} {\bibfnamefont
  {G.}~\bibnamefont {Peña-Castellanos}}, \bibinfo {author} {\bibfnamefont
  {K.}~\bibnamefont {Yamauchi}}, \bibinfo {author} {\bibfnamefont
  {M.}~\bibnamefont {Bussonnier}}, \bibinfo {author} {\bibfnamefont
  {D.}~\bibnamefont {Doncila~Pop}},\ and\ \bibinfo {author} {\bibnamefont
  {et~al.}},\ }\bibfield  {title} {\bibinfo {title} {Napari: a
  multi-dimensional image viewer for python},\ }\bibfield  {journal} {\bibinfo
  {journal} {10.5281/zenodo.6598542}\ }\href
  {https://doi.org/10.5281/zenodo.6598542} {10.5281/zenodo.6598542} (\bibinfo
  {year} {2022})\BibitemShut {NoStop}%
\bibitem [{\citenamefont {Maioli}(2016)}]{maioli2016high}%
  \BibitemOpen
  \bibfield  {author} {\bibinfo {author} {\bibfnamefont {V.~A.}\ \bibnamefont
  {Maioli}},\ }\emph {\bibinfo {title} {High-speed 3-D fluorescence imaging by
  oblique plane microscopy: multi-well plate-reader development, biological
  applications and image analysis}},\ \href@noop {} {Ph.D. thesis},\ \bibinfo
  {school} {Imperial College London} (\bibinfo {year} {2016})\BibitemShut
  {NoStop}%
\bibitem [{\citenamefont {Hoshikawa}\ and\ \citenamefont
  {Saito}(1979)}]{Hoshikawa1979}%
  \BibitemOpen
  \bibfield  {author} {\bibinfo {author} {\bibfnamefont {H.}~\bibnamefont
  {Hoshikawa}}\ and\ \bibinfo {author} {\bibfnamefont {N.}~\bibnamefont
  {Saito}},\ }\bibfield  {title} {\bibinfo {title} {Brownian motion of helical
  flagella},\ }\href {https://doi.org/10.1016/0301-4622(79)80008-3} {\bibfield
  {journal} {\bibinfo  {journal} {Biophysical Chemistry}\ }\textbf {\bibinfo
  {volume} {10}},\ \bibinfo {pages} {81} (\bibinfo {year} {1979})}\BibitemShut
  {NoStop}%
\bibitem [{\citenamefont {Brenner}(1964)}]{Brenner1964}%
  \BibitemOpen
  \bibfield  {author} {\bibinfo {author} {\bibfnamefont {H.}~\bibnamefont
  {Brenner}},\ }\bibfield  {title} {\bibinfo {title} {The stokes resistance of
  an arbitrary particle{\textemdash}{II}},\ }\href
  {https://doi.org/10.1016/0009-2509(64)85051-x} {\bibfield  {journal}
  {\bibinfo  {journal} {Chemical Engineering Science}\ }\textbf {\bibinfo
  {volume} {19}},\ \bibinfo {pages} {599} (\bibinfo {year} {1964})}\BibitemShut
  {NoStop}%
\bibitem [{\citenamefont {Brenner}(1965)}]{brenner1965coupling}%
  \BibitemOpen
  \bibfield  {author} {\bibinfo {author} {\bibfnamefont {H.}~\bibnamefont
  {Brenner}},\ }\bibfield  {title} {\bibinfo {title} {Coupling between the
  translational and rotational brownian motions of rigid particles of arbitrary
  shape i. helicoidally isotropic particles},\ }\href@noop {} {\bibfield
  {journal} {\bibinfo  {journal} {Journal of colloid science}\ }\textbf
  {\bibinfo {volume} {20}},\ \bibinfo {pages} {104} (\bibinfo {year}
  {1965})}\BibitemShut {NoStop}%
\bibitem [{\citenamefont {Cichocki}\ \emph {et~al.}(2015)\citenamefont
  {Cichocki}, \citenamefont {Ekiel-Je{\.{z}}ewska},\ and\ \citenamefont
  {Wajnryb}}]{Cichocki2015}%
  \BibitemOpen
  \bibfield  {author} {\bibinfo {author} {\bibfnamefont {B.}~\bibnamefont
  {Cichocki}}, \bibinfo {author} {\bibfnamefont {M.~L.}\ \bibnamefont
  {Ekiel-Je{\.{z}}ewska}},\ and\ \bibinfo {author} {\bibfnamefont
  {E.}~\bibnamefont {Wajnryb}},\ }\bibfield  {title} {\bibinfo {title}
  {Brownian motion of a particle with arbitrary shape},\ }\href
  {https://doi.org/10.1063/1.4921729} {\bibfield  {journal} {\bibinfo
  {journal} {The Journal of Chemical Physics}\ }\textbf {\bibinfo {volume}
  {142}},\ \bibinfo {pages} {214902} (\bibinfo {year} {2015})}\BibitemShut
  {NoStop}%
\bibitem [{\citenamefont {Gardiner}(2009)}]{Gardiner}%
  \BibitemOpen
  \bibfield  {author} {\bibinfo {author} {\bibfnamefont {C.~W.}\ \bibnamefont
  {Gardiner}},\ }\href@noop {} {\emph {\bibinfo {title} {Handbook of Stochastic
  Methods}}},\ \bibinfo {edition} {3rd}\ ed.\ (\bibinfo  {publisher}
  {Springer-Verlag},\ \bibinfo {address} {Berlin Heidelberg New York},\
  \bibinfo {year} {2009})\ p.\ \bibinfo {pages} {442}\BibitemShut {NoStop}%
\bibitem [{\citenamefont {Arzt}\ \emph {et~al.}(2022)\citenamefont {Arzt},
  \citenamefont {Deschamps}, \citenamefont {Schmied}, \citenamefont {Pietzsch},
  \citenamefont {Schmidt}, \citenamefont {Tomancak}, \citenamefont {Haase},\
  and\ \citenamefont {Jug}}]{arzt2022labkit}%
  \BibitemOpen
  \bibfield  {author} {\bibinfo {author} {\bibfnamefont {M.}~\bibnamefont
  {Arzt}}, \bibinfo {author} {\bibfnamefont {J.}~\bibnamefont {Deschamps}},
  \bibinfo {author} {\bibfnamefont {C.}~\bibnamefont {Schmied}}, \bibinfo
  {author} {\bibfnamefont {T.}~\bibnamefont {Pietzsch}}, \bibinfo {author}
  {\bibfnamefont {D.}~\bibnamefont {Schmidt}}, \bibinfo {author} {\bibfnamefont
  {P.}~\bibnamefont {Tomancak}}, \bibinfo {author} {\bibfnamefont
  {R.}~\bibnamefont {Haase}},\ and\ \bibinfo {author} {\bibfnamefont
  {F.}~\bibnamefont {Jug}},\ }\bibfield  {title} {\bibinfo {title} {Labkit:
  labeling and segmentation toolkit for big image data},\ }\href@noop {}
  {\bibfield  {journal} {\bibinfo  {journal} {Frontiers in Computer Science}\
  }\textbf {\bibinfo {volume} {4}},\ \bibinfo {pages} {777728} (\bibinfo {year}
  {2022})}\BibitemShut {NoStop}%
\bibitem [{\citenamefont {van~der Walt}\ \emph {et~al.}(2014)\citenamefont
  {van~der Walt}, \citenamefont {Schönberger}, \citenamefont {Nunez-Iglesias},
  \citenamefont {Boulogne}, \citenamefont {Warner}, \citenamefont {Yager},
  \citenamefont {Gouillart},\ and\ \citenamefont {Yu}}]{Walt2014}%
  \BibitemOpen
  \bibfield  {author} {\bibinfo {author} {\bibfnamefont {S.}~\bibnamefont
  {van~der Walt}}, \bibinfo {author} {\bibfnamefont {J.~L.}\ \bibnamefont
  {Schönberger}}, \bibinfo {author} {\bibfnamefont {J.}~\bibnamefont
  {Nunez-Iglesias}}, \bibinfo {author} {\bibfnamefont {F.}~\bibnamefont
  {Boulogne}}, \bibinfo {author} {\bibfnamefont {J.~D.}\ \bibnamefont
  {Warner}}, \bibinfo {author} {\bibfnamefont {N.}~\bibnamefont {Yager}},
  \bibinfo {author} {\bibfnamefont {E.}~\bibnamefont {Gouillart}},\ and\
  \bibinfo {author} {\bibfnamefont {T.}~\bibnamefont {Yu}},\ }\bibfield
  {title} {\bibinfo {title} {scikit-image: image processing in python},\ }\href
  {https://doi.org/10.7717/peerj.453} {\bibfield  {journal} {\bibinfo
  {journal} {{PeerJ}}\ }\textbf {\bibinfo {volume} {2}},\ \bibinfo {pages}
  {e453} (\bibinfo {year} {2014})}\BibitemShut {NoStop}%
\bibitem [{\citenamefont {Michalet}(2010)}]{michalet2010mean}%
  \BibitemOpen
  \bibfield  {author} {\bibinfo {author} {\bibfnamefont {X.}~\bibnamefont
  {Michalet}},\ }\bibfield  {title} {\bibinfo {title} {Mean square displacement
  analysis of single-particle trajectories with localization error: Brownian
  motion in an isotropic medium},\ }\href@noop {} {\bibfield  {journal}
  {\bibinfo  {journal} {Physical Review E}\ }\textbf {\bibinfo {volume} {82}},\
  \bibinfo {pages} {041914} (\bibinfo {year} {2010})}\BibitemShut {NoStop}%
\bibitem [{\citenamefont {Yang}\ and\ \citenamefont
  {Bevan}(2017)}]{yang2017interfacial}%
  \BibitemOpen
  \bibfield  {author} {\bibinfo {author} {\bibfnamefont {Y.}~\bibnamefont
  {Yang}}\ and\ \bibinfo {author} {\bibfnamefont {M.~A.}\ \bibnamefont
  {Bevan}},\ }\bibfield  {title} {\bibinfo {title} {Interfacial colloidal rod
  dynamics: Coefficients, simulations, and analysis},\ }\href@noop {}
  {\bibfield  {journal} {\bibinfo  {journal} {The Journal of chemical physics}\
  }\textbf {\bibinfo {volume} {147}},\ \bibinfo {pages} {054902} (\bibinfo
  {year} {2017})}\BibitemShut {NoStop}%
\bibitem [{\citenamefont {Asadi}(2006)}]{asadi2006beet}%
  \BibitemOpen
  \bibfield  {author} {\bibinfo {author} {\bibfnamefont {M.}~\bibnamefont
  {Asadi}},\ }\href@noop {} {\emph {\bibinfo {title} {Beet-sugar handbook}}},\
  \bibinfo {edition} {1st}\ ed.\ (\bibinfo  {publisher} {John Wiley \& Sons},\
  \bibinfo {address} {Hoboken},\ \bibinfo {year} {2006})\BibitemShut {NoStop}%
\bibitem [{\citenamefont {Soesanto}\ and\ \citenamefont
  {Williams}(1981)}]{soesanto1981volumetric}%
  \BibitemOpen
  \bibfield  {author} {\bibinfo {author} {\bibfnamefont {T.}~\bibnamefont
  {Soesanto}}\ and\ \bibinfo {author} {\bibfnamefont {M.~C.}\ \bibnamefont
  {Williams}},\ }\bibfield  {title} {\bibinfo {title} {Volumetric
  interpretation of viscosity for concentrated and dilute sugar solutions},\
  }\href@noop {} {\bibfield  {journal} {\bibinfo  {journal} {The Journal of
  Physical Chemistry}\ }\textbf {\bibinfo {volume} {85}},\ \bibinfo {pages}
  {3338} (\bibinfo {year} {1981})}\BibitemShut {NoStop}%
\bibitem [{\citenamefont {Cox}(1970)}]{cox1970motion}%
  \BibitemOpen
  \bibfield  {author} {\bibinfo {author} {\bibfnamefont {R.}~\bibnamefont
  {Cox}},\ }\bibfield  {title} {\bibinfo {title} {The motion of long slender
  bodies in a viscous fluid part 1. general theory},\ }\href@noop {} {\bibfield
   {journal} {\bibinfo  {journal} {Journal of Fluid mechanics}\ }\textbf
  {\bibinfo {volume} {44}},\ \bibinfo {pages} {791} (\bibinfo {year}
  {1970})}\BibitemShut {NoStop}%
\bibitem [{\citenamefont {Hawkins}\ \emph {et~al.}(2013)\citenamefont
  {Hawkins}, \citenamefont {Sept}, \citenamefont {Mogessie}, \citenamefont
  {Straube},\ and\ \citenamefont {Ross}}]{hawkins2013mechanical}%
  \BibitemOpen
  \bibfield  {author} {\bibinfo {author} {\bibfnamefont {T.~L.}\ \bibnamefont
  {Hawkins}}, \bibinfo {author} {\bibfnamefont {D.}~\bibnamefont {Sept}},
  \bibinfo {author} {\bibfnamefont {B.}~\bibnamefont {Mogessie}}, \bibinfo
  {author} {\bibfnamefont {A.}~\bibnamefont {Straube}},\ and\ \bibinfo {author}
  {\bibfnamefont {J.~L.}\ \bibnamefont {Ross}},\ }\bibfield  {title} {\bibinfo
  {title} {Mechanical properties of doubly stabilized microtubule filaments},\
  }\href@noop {} {\bibfield  {journal} {\bibinfo  {journal} {Biophysical
  journal}\ }\textbf {\bibinfo {volume} {104}},\ \bibinfo {pages} {1517}
  (\bibinfo {year} {2013})}\BibitemShut {NoStop}%
\bibitem [{\citenamefont {Broersma}(1960)}]{Broersma1960}%
  \BibitemOpen
  \bibfield  {author} {\bibinfo {author} {\bibfnamefont {S.}~\bibnamefont
  {Broersma}},\ }\bibfield  {title} {\bibinfo {title} {Viscous force constant
  for a closed cylinder},\ }\href {https://doi.org/10.1063/1.1730995}
  {\bibfield  {journal} {\bibinfo  {journal} {The Journal of Chemical Physics}\
  }\textbf {\bibinfo {volume} {32}},\ \bibinfo {pages} {1632} (\bibinfo {year}
  {1960})}\BibitemShut {NoStop}%
\bibitem [{\citenamefont {Li}\ and\ \citenamefont
  {Tang}(2004)}]{li2004diffusion}%
  \BibitemOpen
  \bibfield  {author} {\bibinfo {author} {\bibfnamefont {G.}~\bibnamefont
  {Li}}\ and\ \bibinfo {author} {\bibfnamefont {J.~X.}\ \bibnamefont {Tang}},\
  }\bibfield  {title} {\bibinfo {title} {Diffusion of actin filaments within a
  thin layer between two walls},\ }\href@noop {} {\bibfield  {journal}
  {\bibinfo  {journal} {Physical Review E}\ }\textbf {\bibinfo {volume} {69}},\
  \bibinfo {pages} {061921} (\bibinfo {year} {2004})}\BibitemShut {NoStop}%
\bibitem [{\citenamefont {Broersma}(1981)}]{broersma1981viscous}%
  \BibitemOpen
  \bibfield  {author} {\bibinfo {author} {\bibfnamefont {S.}~\bibnamefont
  {Broersma}},\ }\bibfield  {title} {\bibinfo {title} {Viscous force and torque
  constants for a cylinder},\ }\href@noop {} {\bibfield  {journal} {\bibinfo
  {journal} {The Journal of Chemical Physics}\ }\textbf {\bibinfo {volume}
  {74}},\ \bibinfo {pages} {6989} (\bibinfo {year} {1981})}\BibitemShut
  {NoStop}%
\end{thebibliography}%

\section*{Materials and methods\label{SI:materials-methods}}

\subsection*{Sample preparation}

\subsubsection*{\emph{E. Coli} flagellum}
Glycerol stock of \emph{E.Coli} strain MG1655WT was grown overnight in \SI{1.5}{\percent} agar with T-broth (\SI{1}{\percent} Tryptone (211705; BD) and \SI{0.5}{\percent} NaCl (S7653; Sigma-Aldrich)) at \SI{37}{\degreeCelsius}. Subsequently, a single colony was taken for inoculation with \SI{10}{\milli \liter} of T-broth in a sterile \SI{125}{\milli \liter} flask. The culture was grown overnight in a rotary shaker, moving at a speed of \num{150}{~RPM} at \SI{30}{\degreeCelsius} until it reached saturation. \SI{100}{\micro \liter} of this culture was diluted in another \SI{10}{\milli \liter} of T-broth in a \SI{125}{\milli \liter} sterile flask and then grown in a rotary shaker moving at a speed of \num{150}{~RPM} at \SI{30}{\degreeCelsius} until it reached \(\text{OD}_{600} = 0.6\). The motility of the bacteria was confirmed using a bright-field microscope. Although the flagella were not visible, the body movement of the bacteria indicated that it had successfully developed long propelling flagella. Subsequently, the bacteria solution was washed three times by centrifugation (\SI{2000}{\g} for \SI{10}{\minute}) at room temperature to separate \emph{E.Coli} from its culture medium. The precipitate was gently resuspended using \SI{10}{\milli \liter} of motility buffer (\SI{10}{\milli \Molar} \(\text{KPO}_4 (P0662; Sigma-Aldrich)\), \SI{67}{\milli \Molar} NaCl (S7653; Sigma-Aldrich), \SI{0.1}{\milli \Molar} EDTA (A15161; Sigma-Aldrich), final pH \num{7.0}). The final suspension was stored at \SI{500}{\micro \liter}.

The amine-coated surface of \emph{E. coli}, including its flagella, was stained with organic dyes tagged with N-Hydroxysuccinimide (NHS)-Ester (PA63101; GE)~\cite{turner2000real}. Since active esters are unstable in moisture, the Cy3B NHS-ester was first prepared in small \SI{3}{\micro \liter} aliquots of \SI{25}{\milli \Molar} in DMSO (85190; ThermoFisher Scientific) and kept in \SI{-20}{\degreeCelsius}. \SI{50}{\micro \liter} of the final bacteria suspension was mixed with a \SI{3}{\micro \liter} aliquot of Cy3B. \SI{2.5}{\micro \liter} of \SI{1}{\Molar} sodium bicarbonate (S6014-500G; Sigma-Aldrich) was added to the new sample to shift its pH to $\sim$\num{7.8}. The sample was then incubated on a slow rotator at room temperature for \SI{1}{\hour}. Subsequently, the sample was washed three times with motility buffer added with \SI[retain-zero-exponent]{e-4}{\percent} Triton X-100 (786-513; G-Biosciences) to remove excess dye and prevent any labeled cells from sticking to the test tube. The washed sample was kept at a volume of \SI{50}{\micro \liter}. The stained flagella were extracted by vortexing and pipetting followed by centrifugation at \SI{8000}{\g} for \SI{5}{\minute} at room temperature. \SI{5}{\micro \liter} of the resulting supernatant was then mixed with \SI{40}{\percent}, \SI{50}{\percent}, and \SI{70}{\percent} (w/v) sucrose (S0389; Sigma-Aldrich) for imaging.

\subsubsection*{Microtubules}
Cycled unlabeled tubulin (032005; PurSolutions) and labeled tubulin-Alexa Fluor 647 (064705; PurSolutions) were mixed in a 4:1 ratio. Single-cycle microtubules were then synthesized by mixing \SI{20}{\micro \Molar} of the tubulin mixture and \SI{1}{\milli \Molar} GMPCPP (NU-405S; Jena Bioscience) in BRB80 buffer (032003; PurSolutions) and \SI{10}{\milli \Molar} DTT (10708984001; Sigma-Aldrich) for \SI{10}{\minute} on ice followed by a \SI{2}{\hour} incubation period at \SI{37}{\degreeCelsius}~\cite{hyman1992role}. This incubation time led to microtubules with an average length of \SI{6.5}{\micro \meter}~\cite{gell2010microtubule}. The sample, diluted to \SI{200}{\nano \Molar}, was then mixed with \SI{90}{\percent} (w/v) sucrose. An additional \SI{10}{\micro \Molar} of Taxol (T7402-5MG; Sigma-Aldrich) was added to stabilize these GMPCPP microtubules. To delay photo-bleaching, the mixture was supplemented with GLOX oxygen scavenging buffer (\SI[per-mode=symbol]{0.02}{\milli \gram \per \milli \liter}, \SI[per-mode=symbol]{0.05}{\milli \gram \per \milli \liter} catalase from bovine liver (C1345-1G; Sigma-Aldrich), \SI[per-mode=symbol]{0.05}{\milli \gram \per \milli \liter} glucose oxidase from \emph{Aspergillus niger} (G2133-10KU; Sigma-Aldrich), and \SI[per-mode=symbol]{3}{\milli \gram \per \milli \liter} glucose (0643-1KG; VWR)).

\subsection*{Imaging chamber preparation}
A \SI{18 x 18}{\milli \meter} coverslip (Thorlabs) and a \SI{24.5 x 76.2}{\milli \meter} microscope slide (VWR) were cleaned with acetone (LC1042044; LabChem) and isopropanol (BDH1133-4LP; VWR), dried, and plasma cleaned using a plasma cleaner (Harrick Plasma) for \SI{10}{\minute}. \SI{100}{\micro \meter} thick double-sided Kapton tape was sandwiched between the coverslip and the microscope slide to create a flow chamber. The chamber was then treated with \SI[per-mode=symbol]{1}{\milli \gram \per \milli \liter} BSA (A4503-10G; Sigma-Aldrich) or \SI[per-mode=symbol]{10}{\milli \gram \per \milli \liter} casein from bovine milk (C7078-500G; Sigma-Aldrich) for \SI{10}{\minute}. \SI{20}{\micro \liter} of the labeled flagella-sucrose sample was flowed into the chamber. After the sample was drained, the flow chamber was sealed by applying epoxy to both chamber openings to prevent evaporation. The epoxy-sealed coverslip was then incubated for an hour at room temperature. 

\subsection*{Oblique plane microscopy}
We modified our previously described stage-scanning high numerical aperture oblique plane microscope (OPM) for high-speed, volumetric, low-inertia acquisition~\cite{sapoznik2020versatile} using a lens-based galvanometric mirror scan/descan unit~\cite{bouchard2015swept,kumar2018integrated,york2019}.  

Briefly, a remote focus microscope consisting of a \num{100}$\times$ NA \num{1.35} silicone oil primary objective (MRD73950; Nikon Instruments), \SI{200}{\milli \meter} primary tube lens (MXA22018; Nikon Instruments), lens-based galvanometric mirror scan/descan unit (\num{2}$\times$ CLS-SL and GVS201; Thorlabs), custom \SI{357}{\milli \meter} secondary tube lens (AC508-500-A and AC508-750-A; Thorlabs)~\cite{york2019}, custom pentaband dichroic (zt405/488/561/640/730rpc-uf3; Chroma Technology Corporation), custom pentaband laser barrier filter (zet405/488/561/640/730m; Chroma Technology Corporation), and \num{40}$\times$ NA \num{0.95} air secondary objective (MRD70470; Nikon Instruments) with a glued anti-reflection coated coverslip (Applied Scientific Instrumentation). A tertiary imaging system imaged the remote image volume, tilted at \ang{30} to the optical axis, consisting of a bespoke solid immersion objective (AMS-AGY v1.0; Special Optics)~\cite{york2019}, custom pentaband laser barrier filter (zet405/488/561/640/730m; Chroma Technology Corporation), \SI{200}{\milli \meter} tertiary tube lens (MXA22018; Nikon Instruments), and camera (OrcaFlash Fusion BT, Hamamatsu Corporation).

Excitation light was provided by a set of \num{5} solid-state lasers (OBIS LX 405–100, OBIS LX 488–150, OBIS LS 561–150, OBIS LX 637–140, and OBIS LX 730–30; Coherent Inc) contained within a control box (Laser Box: OBIS; Coherent Inc). The lasers were steered into a co-linear path using kinematic mirrors and laser combining dichroic mirrors (zt405rdc-UF1, zt488rdc-UF1, zt561rdc-UF1, zt640rdc-UF1; Chroma Technology Corporation). All \num{5} beams were filtered using a \SI{30}{\micro \meter} pinhole spatial filter (P30D; Thorlabs), with the size selected to best filter the \SI{561}{\nano \meter} laser line. After spatial filtering, the beams were expanded and focused to a line by a cylindrical lens (ACY254-075-A; Thorlabs) onto a mirror conjugate to the front focal plane of the primary objective. The line focus was reflected into the optical train of the OPM by the dichroic mirror between the secondary tube lens and the secondary objective. The tilt of the mirror at the line focus controlled translation at the back focal plane of the primary objective and, therefore, the light sheet angle in the sample. The numerical aperture of the light sheet was controlled using an adjustable slit (VA100C; Thorlabs) placed in the back focal plane of the cylindrical lens.

An XYZ positioning stage (FTP-2000; Applied Scientific Instrumentation) and controller (Tiger; Applied Scientific Instrumentation) were used to position the sample into the focal volume of the primary objective.

The microscope was controlled using a Windows 11 Pro 64-bit computer (Thinkstation P620; Lenovo) running custom Python software based around the C++ core of Micromanager via pymmcore-plus and a Napari graphic interface~\cite{edelstein2014advanced,napari2022}. Imaging was deterministically hardware triggered using a programmable digital acquisition device (USB-6341 X Series DAQ; National Instruments). Laser emission was synchronized to the `EXPOSURE OUT' sCMOS signal to eliminate rolling shutter motion blur and galvanometer mirror motion during the camera readout. By scanning the excitation and de-scanning the fluorescence, no inertia was imparted to the sample.

Green fluorescent microspheres (F8803; ThermoFisher) embedded in \SI{1}{\percent} low melting point agarose (R0801; ThermoFisher) were used to calibrate the microscope. The \emph{XYZ} resolution, after deskewing by orthogonal interpolation into the coverslip coordinate system, was \SI[separate-uncertainty = true]{301(21)}{\nm}, \SI[separate-uncertainty = true]{340(19)}{\nm}, \SI[separate-uncertainty = true]{755(18)}{\nm}.

Raw image data were stored as a Zarr file with custom metadata stored as a text file.

\subsection*{Volumetric timelapse acquisition}
The flagella and microtubule solutions were imaged with the \SI{561}{\nano \meter} laser of the OPM setup using a light sheet angle of \ang{30}. The camera region of interest was cropped to \num{1800} px $\times$ \num{256} px and the exposure time was \SI{2}{\milli \second}.  Each volume was acquired using a galvanometric mirror scanning approach~\cite{bouchard2015swept,kumar2018integrated}, where we collected \num{37} images separated by \SI{400}{\nano \meter} at a scanning rate of \SI{0.5}{\kilo \hertz}. This resulted in 3D images over a \SI{180 x 30 x 15}{\micro \meter} parallelepiped shaped volume acquired at a rate of \num{13.3}~volumes/second.

The OPM images were acquired in a coordinate system tilted relative to the coverslip. After the acquisition, each image stack was ``deskewed'' by orthogonal interpolation onto an isotropic 3D grid aligned with the coverslip~\cite{maioli2016high,sapoznik2020versatile}. Optional GPU-accelerated Richardson-Lucy deconvolution (Microvolution) was performed on raw data using the native OPM point spread function for microtubule, but not flagella, experiments.

\subsection*{Numerical simulation of helix diffusion}
Simulation of propulsion matrix parameters using slender body theory~\cite{Johnson1980} and the method of regularized Stokeslets~\cite{Cortez2001} was performed using the Matlab code developed in~\cite{rodenborn2013propulsion} which is available at \url{https://www.mathworks.com/matlabcentral/fileexchange/39265-helical-swimming-simulator}. All parameters must first be normalized to the helix radius, so the simulations were performed with $R = 1$, $L = 30.8$, $\lambda = 10$, and $a = 0.04$.

\subsection*{Non-dimensional propulsion matrix from prior works}
Optical tweezer values~\cite{chattopadhyay2006swimming} have been non-dimensionalized using their parameter estimates $R = \SI{0.21}{\micro \meter}$, $L = \SI{6.5}{\micro \meter}$, and $\theta = \SI{41}{\degree}$. Load cell values~\cite{rodenborn2013propulsion} were taken from their figure 2 using $\theta = \ang{32}$. Gravity values from~\cite{purcell1997efficiency} were taken from the second entry in Table 1, with $\theta = \ang{39}$. Johnson SBT and regularized Stokeslet values were obtained using the code developed in~\cite{rodenborn2013propulsion} with the flagella parameters given in the main text.

\subsection*{Computational resources and data storage}
All computation, except for data acquisition, was performed using a Linux Mint 19 server with 48 computing cores, 1 terabyte of RAM, and two dedicated GPU computing cards (Titan RTX; NVIDIA). All data was stored on a dedicated network attached storage with 780 terabytes of redundant storage (Diskstation DS3018XS with \num{2}$\times$ DX1215 expansion units; Synology). A local fiber network connected the acquisition PC, computational server, and network attached storage.

\vfill


\pagebreak
\clearpage
\setcounter{equation}{0}
\setcounter{figure}{0}

\renewcommand{\theparagraph}{\bf}
\renewcommand{\thesection}{S\arabic{section}}
\renewcommand{\thefigure}{S\arabic{figure}}
\renewcommand{\theequation}{S\arabic{equation}}
\renewcommand{\thealgorithm}{S\arabic{algorithm}}

\onecolumngrid
\flushbottom

\section{Obtaining the propulsion matrix from the fluctuation-\\dissipation theorem and generalized Einstein relations}\label{SI:sec-theory}
Here we develop a simple model of a helix diffusing in a viscous fluid that captures the expected coupling between the helix's translational and rotational motion and use this model to determine the propulsion matrix coefficients in terms of the diffusion coefficients of the system. Although the Stokes equations can capture the full physics of such a situation either with the addition of random thermal forces~\cite{Hoshikawa1979} or using a dissipative Lagrangian approach~\cite{Brenner1964, brenner1965coupling, Brenner1967, Cichocki2015}, we adopt a 1D conservative Lagrangian approach to simplify calculations and obtain physical intuition. In our model, we represent the immersed body as a particle with 1D translational and rotational degrees of freedom and the viscous fluid medium as massive strings coupled to this particle. These massive strings provide effective drag and random forces to the particle in a system that conserves energy, allowing the problem to be solved using a Lagrangian framework.

Using this approach, we will find that the immersed body's velocity and angular velocity are related to the force and torque by a mobility matrix,
\begin{eqnarray}
\begin{pmatrix}
\vec{v}\\
\omega
\end{pmatrix}
&=&
\begin{pmatrix}
\mu_F & \mu_c\\
\mu_c & \mu_\psi
\end{pmatrix}
\begin{pmatrix}
\vec{F} \\
\tau
\end{pmatrix}, \label{eq:ch2-285} 
\end{eqnarray}
which is the inverse of the propulsion matrix $\mu = \vec{P}^{-1}$. This mobility matrix can be used to define a diffusion matrix~\cite{Brenner1964},
\begin{eqnarray}
\vec{D} &=& k_B T \mu \label{eq:generalized_einstein}\\
\vec{P} &=& k_B T \vec{D}^{-1}, 
\end{eqnarray}
\noindent where $k_B$ is the Boltzmann constant and $T$ is the absolute temperature.

We will show that the system diffuses according to this matrix's coefficients. Eq.~\ref{eq:generalized_einstein} is a generalization of the Einstein relation, which gives the diffusion coefficient $D = k_b T \mu$ in terms of the mobility $\mu$ defined by $v = \mu F$ for a system with one degree of freedom~\cite{einstein1905erzeugung}. In contrast, eq.~\ref{eq:generalized_einstein} describes a coupled system with two degrees of freedom. Eq.~\ref{eq:generalized_einstein} implies that the propulsion matrix can be determined from diffusion measurements, as demonstrated in the main text.

\subsection{A freely diffusing system with a translation degree of freedom\label{sec:1D-particle}}
Consider a particle representing the helix or other immersed body, coupled to a string, representing the effect of the fluid on the helix. Although this combined system is conservative, the particle can dissipate energy by propagating waves down to the string. Similarly, incoming waves along the string will drive the particle. If these incoming waves have a thermal distribution, they will induce Brownian fluctuations of the particle and serve as a heat bath model.

We begin by writing the Lagrangian for this system and obtaining the equations of motion. Suppose the particle has mass $M$ and is restricted to lie along $x=0$. The string extends along $x>0$ at height $Y(x, t)$ and has a mass per unit length $\rho$ and line tension of $\sigma$. The Lagrangian density of the system is given by
\begin{eqnarray}
    {\cal L} & = & \delta(x) \frac{M}{2} \left( \frac{\partial Y}{\partial t} \right)^2
    + \ u_+(x) \left[ \frac{\rho}{2} \left( \frac{\partial Y}{\partial t} \right)^2
    - \frac{\sigma}{2} \left( \frac{\partial Y}{\partial x} \right)^2 \right],
    \label{eq:ch2-1}
\end{eqnarray}
where $\delta(x)$ is the Dirac delta function and $u_+(x)$ is the Heaviside step function. Note that  $Y(0,t)$ is the displacement of the particle along $y$. The quantity in square brackets is the Lagrangian density of a string under tension.

The equations of motion of this system are given by the Euler-Lagrange equations corresponding to eq.~(\ref{eq:ch2-1}), which are
\begin{eqnarray}
    M \frac{d^2Y(0,t)}{dt^2} - \left. \sigma \frac{\partial Y}{\partial x} \right\vert_{x = 0^+} &=& 0 \label{eq:ch2-12}\\
    \rho \frac{\partial^2 Y}{\partial t^2} - \sigma \frac{\partial^2 Y}{\partial x^2} &=& 0. \label{eq:ch2-6}
\end{eqnarray}
Equation~\ref{eq:ch2-12} describes the motion of the particle, and the second term is the force the string exerts on the particle. Equation~\ref{eq:ch2-6} is the wave equation for the string.

From the perspective of the particle, the string coupling acts as either a damping or driving force. Since Eq.~(\ref{eq:ch2-12}) is a wave equation, we can write the solutions as a sum of incoming (left-going) and outgoing (right-going) waves
\begin{equation}
    Y(x,t) = Y_\text{in}(t + x/v) + Y_\text{out}(t - x/v)
    \label{eq:ch2-13}
\end{equation}
where $Y_\text{in}$ and $Y_\text{out}$ are arbitrary functions with $v \equiv \sqrt{\sigma/\rho}$ the propagation velocity of waves along the string. Physically, incoming waves represent driving forces exerted on the particle, while outgoing waves represent damping forces.

Taking the derivative of eq.~(\ref{eq:ch2-13}) at $x = 0^+$ and substituting into eq.~(\ref{eq:ch2-12}) yields
\begin{eqnarray}
    M \frac{d^2 Y}{dt^2} &=& - \Gamma \frac{dY}{dt} + F(t) \label{eq:ch2-25}\\
    F(t) &\equiv& 2\Gamma \frac{dY_\text{in}}{dt}, \label{eq:ch2-24}
\end{eqnarray}
where $\Gamma \equiv \sigma/v = \sqrt{\rho/\sigma}$ is the wave impedance of the string. The first term on the right-hand side of eq.~\ref{eq:ch2-25} represents a damping or drag force where the moving particle transfers energy to propagating modes on the string. The second term is the driving force that the string exerts on the particle. Although eq.~\ref{eq:ch2-25} has a dissipative element, the system as a whole does not dissipate energy.

\subsection{Free diffusion in a viscous medium}
In the analogy between our particle-string system and a helix immersed in a viscous medium, the drag force on the particle corresponds to the viscous damping force that the medium exerts on the helix. To model a free particle in a viscous medium at a low Reynolds number, we suppose the inertial term in eq.~\ref{eq:ch2-25} can be neglected compared with the dissipative term, leading to
\begin{equation}
    \Gamma \frac{d Y}{dt} = F. \label{eq:ch2-78}
\end{equation}
To model the diffusion of the helix in a viscous medium that serves as a heat bath, we suppose that $F$ is a fluctuating force, making eq.~\ref{eq:ch2-78} a Langevin equation~\cite{Gardiner}. We suppose $F$ has white-noise correlation statistics,
\begin{eqnarray}
    \langle F(t) \rangle &=& 0 \label{eq:ch2-58}\\
     \langle F(t) F(t') \rangle &=& 2 \Gamma k_B T \ \delta(t - t'). \label{eq:ch2-77}
\end{eqnarray}
The value of the delta function prefactor is required to obtain correct equilibrium behavior, i.e., to satisfy the equipartition theorem. Equation~\ref{eq:ch2-77} indicates that the correlations of the fluctuating force are proportional to the damping constant $\Gamma$ and the characteristic thermal energy $k_B T$. This is an example of what Kubo referred to as the second fluctuation-dissipation theorem~\cite{kubo1966fluctuation}.

We now demonstrate that the particle undergoes Brownian diffusion by showing that the mean-square displacement takes the form $2Dt$ where $D$ is the diffusion coefficient. Integrating eq.~\ref{eq:ch2-78} and taking the ensemble average, we find
\begin{eqnarray}
    \langle [Y(t) - Y(0)]^2 \rangle &=&
    \frac{1}{\Gamma^2} \int_0^t \int_0^t \langle F(t')F(t'') \rangle dt'' dt' = \frac{2k_BT}{\Gamma} t\label{eq:ch2-82},
\end{eqnarray}
or $D = \frac{k_BT}{\Gamma}$, which provides a connection between the viscous drag coefficient and diffusion.

\subsection{A freely diffusing system with dynamically coupled translation and rotation}
The simplest way to couple translation and rotation is to introduce a term of the form $Y \dot{\psi}$ into the Lagrangian. However, it is straightforward to show that such a term leads to an asymmetric ``propulsion matrix'' and does not capture the desired phenomenology. Instead, we introduce a new dynamic variable, $Z$, to capture the propulsion matrix physics, which couples the translational and rotational motion. This is motivated by the idea that the fluid physically mediates translational-rotational coupling. We also introduce an additional string-like heat bath coupled to $Z$. Our Lagrangian density now becomes
\begin{eqnarray}
    \mathcal{L} & = & \delta(x) \left\{ \frac{M}{2} \left( \frac{\partial Y}{\partial t} \right)^2
    + C_1 \frac{\partial Y}{\partial t} Z \right\}  + \ u_+(x) \left[ \frac{\rho}{2} \left( \frac{\partial Y}{\partial t} \right)^2
    - \frac{\sigma}{2} \left( \frac{\partial Y}{\partial x} \right)^2 \right] \notag \\
    & & 
    + \delta(x) \left\{ \frac{I}{2} \left( \frac{\partial \psi}{\partial t} \right)^2
    + C_2 \frac{\partial \psi}{\partial t}Z \right\} + \ u_+(x) \left[ \frac{\eta}{2} \left( \frac{\partial \psi}{\partial t} \right)^2
    - \frac{\xi}{2} \left( \frac{\partial \psi}{\partial t} \right)^2 \right] \notag \\
    & &
    + \ u_+(x) \left[ \frac{\rho_Z}{2} \left( \frac{\partial Z}{\partial t} \right)^2
    - \frac{\sigma_Z}{2} \left(\frac{\partial Z}{\partial x} \right)^2 \right], \label{eq:ch2-181} 
\end{eqnarray}
where $C_1$ and $C_2$ are coupling constants parameterizing the interaction between $Y$ and $Z$ or $\psi$ and $Z$, respectively. We have introduced the heat bath coupled to $Z$ in the third line. It should be kept in mind that all heat baths are equivalent; hence, modeling the heat bath aspect of the fluid by massive strings is done as a matter of convenience.

Since there are three fields, there are now three Euler-Lagrange equations with variables $Y$, $\psi$, and $Z$. Using the same techniques as in the previous sections and neglecting the inertial terms consistent with the low Reynolds number limit results in
\begin{eqnarray}
    \Gamma \frac{dY}{dt} + C_1 \frac{d Z}{dt} &=& F \label{eq:ch2-192} \\
    C_2 \frac{dZ}{dt} + \Gamma_\psi \frac{d \psi}{dt}  &=& T_\psi \label{eq:ch2-193} \\
    - C_1 \frac{dY}{dt} - C_2 \frac{d \psi}{dt} + \Gamma_Z \frac{d Z}{dt} &=& F_Z, \label{eq:ch2-194}
\end{eqnarray}
where $\Gamma \equiv \sqrt{\rho/\sigma}$, $\Gamma_\psi \equiv \sqrt{\eta/\xi}$, and $\Gamma_Z \equiv \sqrt{\rho_Z/\sigma_Z}$.
By rewriting these as expressions for $F$ and $T_\psi$ and eliminating $dZ/dt$ we can put these equations in a similar form to the propulsion matrix,
\begin{align}
    F &= \left( \Gamma + \frac{C_1^2}{\Gamma_Z} \right) \frac{dY}{dt}
    + \frac{C_1 C_2}{\Gamma_Z} \frac{d\psi}{dt}
    + \frac{C_1}{\Gamma_Z} F_Z 
    \label{eq:ch2-200}\\
    T_\psi &= \frac{C_1 C_2}{\Gamma_Z} \frac{dY}{dt}
    + \left( \Gamma_\psi + \frac{C_2^2}{\Gamma_Z} \right) \frac{d\psi}{dt} 
    + \frac{C_2}{\Gamma_Z} F_Z,
    \label{eq:ch2-201}
\end{align}
which implies that the propulsion matrix coefficients are
\begin{eqnarray}
    A & = & \Gamma + \frac{C_1^2}{\Gamma_Z} \label{eq:ch2-202} \\
    B & = & \frac{C_1C_2}{\Gamma_Z} \label{eq:ch2-203} \\
    D & = & \Gamma_\psi + \frac{C_2^2}{\Gamma_Z}, \label{eq:ch2-204}
\end{eqnarray}
since in this case we have $v = dY/dt$ and $\omega = d\psi/dt$.

The constants $C_1$ and $C_2$ can be eliminated from eqs.~(\ref{eq:ch2-200}) and (\ref{eq:ch2-201}) using eqs.~\ref{eq:ch2-202}--\ref{eq:ch2-204} which yields
\begin{align}
    F &= A \frac{dY}{dt} + B \frac{d\psi}{dt}
    + \sqrt{A-\Gamma} \frac{F_Z}{\sqrt{\Gamma_Z}}
    \label{eq:ch2-215}\\
    T_\psi &= B \frac{dY}{dt}
    + D \frac{d\psi}{dt} 
    + \sqrt{D - \Gamma_\psi} \frac{F_Z}{\sqrt{\Gamma_Z}}.
    \label{eq:ch2-216}
\end{align}

Next, we invert Eqs.~\ref{eq:ch2-215} and \ref{eq:ch2-216} to obtain the velocity and angular velocity as functions of the forces and torques,
\begin{align}
    \frac{dY}{dt} &= \left(\frac{D}{AD - B^2} \right)F - \left(\frac{B}{AD - B^2}\right) T_\psi
    - \left(\frac{D\sqrt{A - \Gamma} - B\sqrt{D - \Gamma_\psi}}
    {(AD - B^2) \sqrt{\Gamma_z}}\right) F_Z
    \label{eq:ch2-219}\\
    \frac{d\psi}{dt} &= - \left(\frac{B}{AD - B^2}\right)F + \left(\frac{A}{AD - B^2}\right) T_\psi
    + \left(\frac{B\sqrt{A - \Gamma} - A \sqrt{D - \Gamma_\psi}}
    {(AD - B^2) \sqrt{\Gamma_z}} \right) F_Z.
    \label{eq:ch2-220}
\end{align}
These expressions have the same form as eq.~\ref{eq:ch2-285}, and therefore we identify the coefficients of $F$ and $T_\psi$ as the mobilities discussed previously. As mentioned before and shown explicitly here, the mobilities are the elements of the inverse of the propulsion matrix.

Equations~\ref{eq:ch2-219}--\ref{eq:ch2-220} are a pair of coupled Langevin equations that can be solved using well-known techniques once the force correlation functions are known~\cite{Gardiner}. As such, we expect diffusive motion of our $Y$ and $\psi$ degrees of freedom at long times, and we wish to determine the diffusion coefficients in terms of the propulsion matrix coefficients $A$, $B$, and $D$. We assume that the force-force and torque-torque correlations are given by
\begin{eqnarray}
 \langle F(t) F(t') \rangle &=& 2 \Gamma k_B T \ \delta(t - t') \label{eq:force_force_corr}\\
 \langle T_\psi(t) T_\psi(t') \rangle &=& 2 \Gamma_\psi k_B T \ \delta(t - t') \label{eq:torque_torque_corr}\\
 \langle F_Z(t) F_Z(t') \rangle &=& 2 \Gamma_Z k_B T \ \delta(t - t') \label{eq:fz_fz_corr},
\end{eqnarray}
and that cross force-force correlations are zero. Compared to~\cite{Hoshikawa1979}, the extra dynamic degree of freedom $Z$ avoids the need to assume non-zero torque-force correlations.

We obtain expressions for the mean-square displacement, the mean-square angular displacement, and the mixed displacement by substituting eqs.~\ref{eq:force_force_corr}--\ref{eq:fz_fz_corr} into eqs.~\ref{eq:ch2-219}--\ref{eq:ch2-220} and integrating. We find the translational diffusion coefficient $D_\parallel$, the rotational diffusion coefficient $D_\psi$, and the codiffusion coefficient $D_{\parallel\psi}$ are
\begin{eqnarray}
\begin{pmatrix}
D_\parallel & D_{\parallel\psi}\\
D_{\parallel\psi} & D_\psi
\end{pmatrix}
&=&
\frac{1}{AD - B^2}
\begin{pmatrix}
D & -B\\
-B & A
\end{pmatrix}
k_b T = \vec{P}^{-1} k_b T
\label{eq:dmat_from_pmat}
\end{eqnarray}
which proves eq.~\ref{eq:generalized_einstein}.

Inverting eq.~\ref{eq:dmat_from_pmat}, we obtain the desired expression for the propulsion matrix coefficients in terms of the diffusion coefficients
\begin{eqnarray}
\begin{pmatrix}
    A & B\\
    B & D
\end{pmatrix}
&=&
\frac{1}{D_\parallel D_\psi - D_{\parallel\psi}^2}
\begin{pmatrix}
    D_\parallel & -D_{\parallel\psi}\\
    -D_{\parallel\psi} & D_\psi
\end{pmatrix}
k_B T = \vec{D}^{-1} k_b T
\end{eqnarray}

\section{Propulsion Matrix with Helical Symmetry\label{SI:sec-symmetry}}
For a generic body, the propulsion matrix is a $6 \times 6$ symmetric matrix that relates the 6 vector components of forces and torques to the 6 vector components of velocities and angular velocities. As the matrix is symmetric, it has only \num{21} independent components. The matrix is often divided into $3 \times 3$ submatrices $K_T$, $K_R$, and $K_C$ as in eq.~1 of the main text.
When a body has symmetry, the number of independent components is further reduced~\cite{Brenner1964}. Note that $K_T$ is always symmetric and independent of the choice of coordinate origin. $K_R$ is symmetric but dependent on the choice of origin. $K_C$ is not necessarily symmetric and also dependent on the choice of origin. However, $K_C$ is symmetric about one special origin, called the center of reaction.

Here, we consider how helical symmetry affects the propulsion matrix. We consider a helix with an axis directed along $\uvec{z}$ and a center line parameterized by
\begin{eqnarray}
\vec{R}(z) &=& r \cos \left( \frac{2 \pi}{p} z \right) \uvec{x} + r \sin \left(\frac{2 \pi}{p} z \right) \uvec{y} + z \uvec{z},
\quad \quad z \in \left[-\frac{L}{2}, \frac{L}{2}\right],
\label{eq:helix_center_line}
\end{eqnarray}
where $p$ is the helical pitch and $r$ is the helical radius. If $p<0$ the helix is left-handed, while if $p>0$ it is right-handed. Comparing these coordinates with the body axes, we have $\uvec{n}_1 = \uvec{z}$, $\uvec{n}_2 = \uvec{x}$, $\uvec{n}_3 = \uvec{y}$.

An infinitely long helix possess two types of symmetry: (1) discrete \SI{180}{\degree} rotation symmetry along the $\uvec{x}$ axis and (2) screw symmetry which can be described by first rotating about $\uvec{z}$ by arbitrary angle $\theta$ and then translating along $\uvec{z}$ by distance $\frac{\theta}{2\pi} p$. A finite helix only has symmetry (1).

Brenner explicitly considers discrete \SI{180}{\degree} rotation symmetry along the $\uvec{x}$ axis in~\cite{Brenner1964}, and finds $K_T$, $K_C$, and $K_R$ all have the form
\begin{eqnarray}
K &=&
\begin{pmatrix}
K_{11} & 0 & 0\\
0 & K_{22} & K_{23}\\
0 & K_{32} & K_{33}
\end{pmatrix},
\end{eqnarray}
where $\left(K_T \right)_{23} = \left(K_T\right)_{32}$ and $\left(K_R\right)_{23} = \left(K_R\right)_{32}$, but $\left(K_C\right)_{23}$ and $\left(K_C\right)_{32}$ may differ.

The screw symmetry is an affine transformation involving translation and rotation. Brenner only considers point-symmetries in~\cite{Brenner1964}, but his approach can easily be generalized to account for affine transformations using his transformation rules for $K^C$ and $K^R$ when the origin shifts. However, in this case, developing this generalization is unnecessary because the rotational and translational symmetry operations commute. This implies that the correction terms for $K_C$ and $K_R$ cancel, and the screw symmetry acts the same as rotational symmetry about the $z$-axis. Combining this symmetry with the discrete $\uvec{x}$-rotation symmetry yields,
\begin{eqnarray}
K_T &=&
\begin{pmatrix}
K^T_{11} & 0 & 0\\
0 & K^T_{11} & 0\\
0 & 0 & A
\end{pmatrix}\\
K_C &=&
\begin{pmatrix}
K^C_{11} & 0 & 0\\
0 & K^C_{11} & 0\\
0 & 0 & B
\end{pmatrix}\\
K_R &=&
\begin{pmatrix}
K^R_{11} & 0 & 0\\
0 & K^R_{11} & 0\\
0 & 0 & D
\end{pmatrix}.
\end{eqnarray}
Therefore for the infinite helix, there are only \num{6} unique components of the propulsion matrix, and all submatrices decouple, so we may rewrite this as a set of  $2 \times 2$ form of the propulsion matrix about the helical axes as in eq.~2 of the main text. Although this argument strictly holds only for the infinite helix, we expect the propulsion matrix for the finite helix to approach that of the infinitely long helix as $L \to \infty$. 

We expect $K^C_{11} = 0$, but this is a property of the helix shape and cannot be deduced from symmetry.

\section{3D tracking of flagella\label{SI:sec-data-analysis}}

\subsection{3D helix segmentation\label{SI:sec-segmentation}}
Starting from the deskewed time-series volumetric image data, we first manually identified individual flagella and selected a region of interest (ROI) of size $150~{\rm px}\times150~{\rm px}\times128~{\rm px}$ along the $x$, $y$, and $z$ axes. Only data sets where the flagellum stayed in this ROI for at least \num{100} time points were analyzed. We segmented the image pixels for each region of interest into the foreground (i.e., part of the flagellum) and background using LabKit, a random forest-based pixel classification algorithm~\cite{arzt2022labkit}. When more than a single flagellum was present per ROI, we isolated the flagellum of interest by identifying clusters using the \texttt{label} function from the scikit-image \texttt{measure} submodule and retaining only the largest cluster~\cite{Walt2014}. We extracted the coordinates of the foreground pixels from the segmented images, which served as the basis for further analysis.

\subsection{Tracking the helix position and orientation}
The center-of-mass position ($\vec{R}$) and three unit vectors, or body axes (fixed relative to the helix), ($\uvec{n}_1, \uvec{n}_2, \uvec{n}_3$), parameterize the position and orientation of a helix (see Fig.~2c of the main text). $\uvec{n}_1$ is the unit vector parallel to the long axis of the helix that passes through the center of mass, $\uvec{n}_2$ as the unit vector orthogonal to $\uvec{n}_1$ which passes through both the center of mass and a point along the center line of the helix, and $\uvec{n}_3 = \uvec{n}_1 \times \uvec{n}_2$. Identification of the body axes is essential since the helix tumbles as it diffuses and the propulsion matrix coefficients depend on the choice of coordinates. Therefore, to properly evaluate helix diffusion, we tracked the translations and rotations of a flagellum with respect to its body axes.

\begin{figure*}[h!]
\centering
\includegraphics[width=1.0\linewidth]{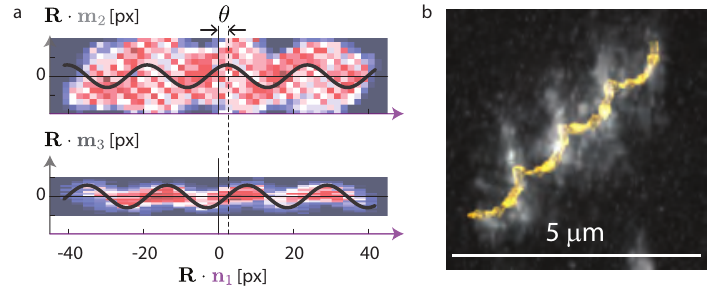}
\caption{
    \textbf{Determination of the flagella position and orientation.}
    ({\bf a}) Helix foreground coordinates projected onto the $\uvec{n}_1$--$\uvec{m}_2$ and $\uvec{n}_1$--$\uvec{m}_3$ planes with the center of mass taken as the origin. The foreground points are shown as 2D histograms. The black line is a simultaneous fit to eqs.~38-39.
    ({\bf b}) Isometric view of the deskewed fluorescence image overlapped with left-handed helix center line obtained from the curve fit (orange).
\label{fig:SI-fitting}}
\end{figure*}

We determined the center-of-mass coordinate using the coordinates of the foreground voxels ($\vec{R}_i$),
\begin{equation}
    \vec{R} = \frac{1}{N} \sum_{i=1}^N \vec{R}_i \label{eq:DA-cm}
\end{equation}
where $N$ is the total number of foreground voxels.

We estimated $\uvec{n}_1$ by performing a principal component analysis (PCA) on the foreground coordinates and selecting the most significant principal component. Due to symmetry, $\uvec{n}_1$ is only determined up to a factor of $\pm 1$. Therefore, we chose the sign of $\uvec{n}_1$ by minimizing the angular difference with the previous frame.

To determine $\uvec{n}_2$ and $\uvec{n}_3$, we started by assigning an arbitrary unit vector $\uvec{m}_2$ orthogonal to $\uvec{n}_1$ and $\uvec{m}_3 = \uvec{n}_1 \times \uvec{m}_2$. Next, we projected the helix coordinates onto the $\uvec{n}_1$--$\uvec{m}_2$ and $\uvec{n}_1$--$\uvec{m}_3$ planes after taking the center of mass as the origin. Considering the helical shape of the flagella (eq.~\ref{eq:helix_center_line}), the expected profiles are
\begin{eqnarray}
\left(\vec{R}_i - \vec{R} \right) \cdot \uvec{n}_2 &=& r \cos \left( \frac{2\pi}{p} \left(\vec{R}_i - \vec{R} \right) \cdot \uvec{n}_1 + \theta \right) \label{eq:helix_r2}\\
\left(\vec{R}_i - \vec{R} \right) \cdot \uvec{n}_3 &=& r \sin \left( \frac{2\pi}{p} \left(\vec{R}_i - \vec{R} \right) \cdot \uvec{n}_1 + \theta \right) \label{eq:helix_r3},
\end{eqnarray}
where the phase $\theta \neq 0$ measures how far $\uvec{m}_2$ is from $\uvec{n}_2$. We determined $\theta$ with a simultaneous nonlinear least-squares fit of the projected coordinates to eqs.~\ref{eq:helix_r2} and \ref{eq:helix_r3} (Fig.~\ref{fig:SI-fitting}) and then computed
\begin{equation}
    \uvec{n}_2 = \uvec{m}_2 \cos{\theta} - \uvec{m}_3 \sin{\theta} \label{eq:DA-3}.
\end{equation}

The unit vectors $\uvec{n}_1(t)$, $\uvec{n}_2(t)$, and $\uvec{n}_3(t)$ provide an intuitive parameterization of the orientation of the helix. However, rotational diffusion is more conveniently described by angular displacements. Therefore, we would like to determine the unit vectors' instantaneous rotation angles about the body axes. Note that, unlike a description of orientation in terms of Euler angles that describe rotation relative to a fixed set of axes, we are interested in the rotation angles about the axes at time $t$, which generate a new set of axes at time $t + dt$. We will refer to rotations about $\uvec{n}_1(t)$, $\uvec{n}_2(t)$ and $\uvec{n}_3(t)$ through angles $\psi_1$, $\psi_2$, and $\psi_3$ as roll, yaw, and pitch, respectively.

For a pure rotation about $\uvec{n}_1$ with displacement $\Delta \psi_1$,
\begin{eqnarray}
\uvec{n}_1(t + dt) & = & \uvec{n}_1(t) \label{eq:3-3} \\
\uvec{n}_2(t + dt) & = & \cos(\Delta \psi_1) \uvec{n}_2(t) + \sin(\Delta \psi_1) \uvec{n}_3(t) \label{eq:3-4} \\
\uvec{n}_3(t + dt) & = & - \sin(\Delta \psi_1) \uvec{n}_2(t) + \cos(\Delta \psi_1) \uvec{n}_3(t) \label{eq:3-5}
\end{eqnarray}
Taylor expanding the trigonometric functions to linear order, we obtain
\begin{equation}
\frac{\partial \uvec{n}_1}{\partial \psi_1} =  0, \quad
\frac{\partial \uvec{n}_2}{\partial \psi_1} =  \uvec{n}_3, \quad
\frac{\partial \uvec{n}_3}{\partial \psi_1}  = -\uvec{n}_2 . \label{eq:3-14}
\end{equation} 
Similarly, for a pure pitch or yaw,
\begin{alignat}{3}
&\frac{\partial \uvec{n}_1}{\partial \psi_2} &= \uvec{n}_2, &\quad
\frac{\partial \uvec{n}_2}{\partial \psi_2} = -\uvec{n}_1, &\quad
&\frac{\partial \uvec{n}_3}{\partial \psi_2} = 0 \label{eq:3-35}\\
&\frac{\partial \uvec{n}_1}{\partial \psi_3} &= -\uvec{n}_3, &\quad
\frac{\partial \uvec{n}_2}{\partial \psi_3} = 0, &\quad
&\frac{\partial \uvec{n}_3}{\partial \psi_3} = \uvec{n}_1.  \label{eq:3-23}
\end{alignat}

We would like to find expressions for the time derivatives of the angles, which we can do by substituting eqs.~\ref{eq:3-14}--\ref{eq:3-35} into the chain rule expression
\begin{eqnarray}
\frac{d \uvec{n}_i}{dt} &=& \sum_{j=1}^3 \frac{\partial \uvec{n}_i}{\partial \psi_j} \frac{d \psi_j}{dt}.  \label{eq:3-36}
\end{eqnarray}
Solving eq.~\ref{eq:3-36} for the angular time derivatives gives
\begin{align}
\frac{d\psi_1}{dt} &=  \uvec{n}_3 \cdot \frac{ d \uvec{n}_2}{dt} \label{eq:3-51}\\
\frac{d\psi_2}{dt} &=  \uvec{n}_2 \cdot \frac{d \uvec{n}_1}{dt} \label{eq:3-53}\\
\frac{d\psi_3}{dt} &=  \uvec{n}_1 \cdot \frac{d \uvec{n}_3}{dt}. \label{eq:3-52}
\end{align} 
The angular displacements ($\Delta_{\psi_i}$) can be determined by integration.

To estimate the translational displacement of the helix along the body axes, we projected the center-of-mass displacement. For a given time interval $\tau$, the translational displacement along body axis $\uvec{n}_i$ is 
\begin{equation}
    \Delta_{n_i}(\tau) = \int_0^\tau \frac{\partial \vec{R}}{\partial t} \cdot \uvec{n}_i(t) \, {\rm dt}.  \label{eq:3-54}
\end{equation}

\begin{algorithm}[h!]
\caption{MSD evaluation for translation along the body axes} \label{alg:SI-MSD-trans}
\begin{algorithmic}[1]
\Require center of mass locations $\vec{R}[j]$ at each frame $j=0,...N - 1$
\Require unit vectors $\uvec{n}_1[j], \uvec{n}_2[j], \uvec{n}_3[j]$ at each frame
\Require maximum time-lag $\tau_\text{max}$
\Ensure MSD along the body axes for lag times $1, ... \tau_\text{max}$
\State $j \leftarrow 1$
\While{$j < \tau_\text{max}$} \Comment{Loop over time-lags}
	\State $\Delta_{n_1}$, $\Delta_{n_2}$, $\Delta_{n_3}$ $\leftarrow$ [ ]
	\State $i \leftarrow 0$
	 \While{$i+j < N $} \Comment{Loop over frames in trajectory}
        \State SubStep$_{n_1}$, SubStep$_{n_2}$, SubStep$_{n_3}$ $\leftarrow$ [ ]
		\State $k \leftarrow 0$
		 \While{$k < j$} \Comment{Finite difference integration of eq.~\ref{eq:3-54}}
		    \State $\Delta \vec{R} \leftarrow {\vec{R}[i+k+1]-\vec{R}[i+k]}$
		    \State SubStep$_{n_1}$.append $\leftarrow$ $\uvec{n}_1[i+k]\cdot\Delta\vec{R}$
		    \State SubStep$_{n_2}$.append $\leftarrow$ $\uvec{n}_2[i+k]\cdot\Delta\vec{R}$
		    \State SubStep$_{n_3}$.append $\leftarrow$ $\uvec{n}_3[i+k]\cdot\Delta\vec{R}$
		    \State $k \leftarrow k + 1$
	   \EndWhile
		 \State $\Delta_{n_1}$.append $\leftarrow$ Sum(SubStep$_{n_1}$)
		 \State $\Delta_{n_2}$.append $\leftarrow$ Sum(SubStep$_{n_2}$)
		 \State $\Delta_{n_3}$.append $\leftarrow$ Sum(SubStep$_{n_3}$)
   	 \State $i \leftarrow i + 1$
	 \EndWhile
	 \State MSD$_{n_1}[j-1]$ $\leftarrow$ Mean($\Delta_{n_1}^2$)
	 \State MSD$_{n_2}[j-1]$ $\leftarrow$ Mean($\Delta_{n_2}^2$)
  \State MSD$_{n_3}[j-1]$ $\leftarrow$ Mean($\Delta_{n_3}^2$)
	 \State $j \leftarrow j + 1$
\EndWhile
\end{algorithmic}
\end{algorithm}

In our numerical implementation, we worked with a left-handed coordinate system, which is natural because conventionally, the top of the image coincides with the lowest $y$-coordinates. In these coordinates, we used a right-handed helix model ($p>0$ in eqs.~\ref{eq:helix_r2}--\ref{eq:helix_r3}). When the space is reflected in $y$, this results in a left-handed helix, as expected for the \emph{E. coli} flagella. This choice does not impact the determination of any of the diffusion coefficients because all step-size quantities (eqs.~\ref{eq:3-51}--\ref{eq:3-54}) are computed from the inner products of vectors, which are invariant under reflection.

\subsection{Mean squared displacement\label{SI:sec-msd}}

After computing the helix displacement along ($\Delta_{n_i}$) and the rotational displacement about ($\Delta_{\psi_i}$) the body axes using eqs.~\ref{eq:3-51}--\ref{eq:3-54}, we estimated the diffusion coefficients from
\begin{equation}
\langle\Delta_\xi(\delta t) \Delta_\eta(\delta t) \rangle = 2 D_{\xi, \eta} \delta t + C_{\xi, \eta} \label{eq:msd_fit}
\end{equation}
for $\xi, \eta = n_1, n_2, n_3, \psi_1, \psi_2, \psi_3$. When $\eta = \xi$ these are true translational or rotational diffusion coefficients, the correlation function is the mean squared displacement (MSD) or the mean squared angular displacement (MSAD). Otherwise, they represent correlations between steps along different coordinates. The correlations were computed from the displacements using Alg.~\ref{alg:SI-MSD-trans}. Unlike a spherical particle, the translation displacement $\Delta_{n_i}$ must be computed by integrating a finite-difference approximation of eq.~\ref{eq:3-54}.

To determine the diffusion coefficients from the MSD values, we evaluated the slope by fitting the first few time lags to a linear function. The optimal number of fitting points depends on the interplay of localization error and statistical uncertainty. In the absence of a localization error, the optimal method is to use only the first two points. When localization error is present, using more points produces a more accurate result~\cite{michalet2010mean}. On the other hand, the number of points from the trajectory used to generate MSD points versus time-lags decreases linearly, leading to large statistical uncertainty for long time-lags. In this work, we extract the slope from the first ten points for all correlation functions, which appears to be a good compromise.

\subsection{Verification with synthetic data}

\begin{figure}[ht!]
\centering
\includegraphics[width=1.0\textwidth]{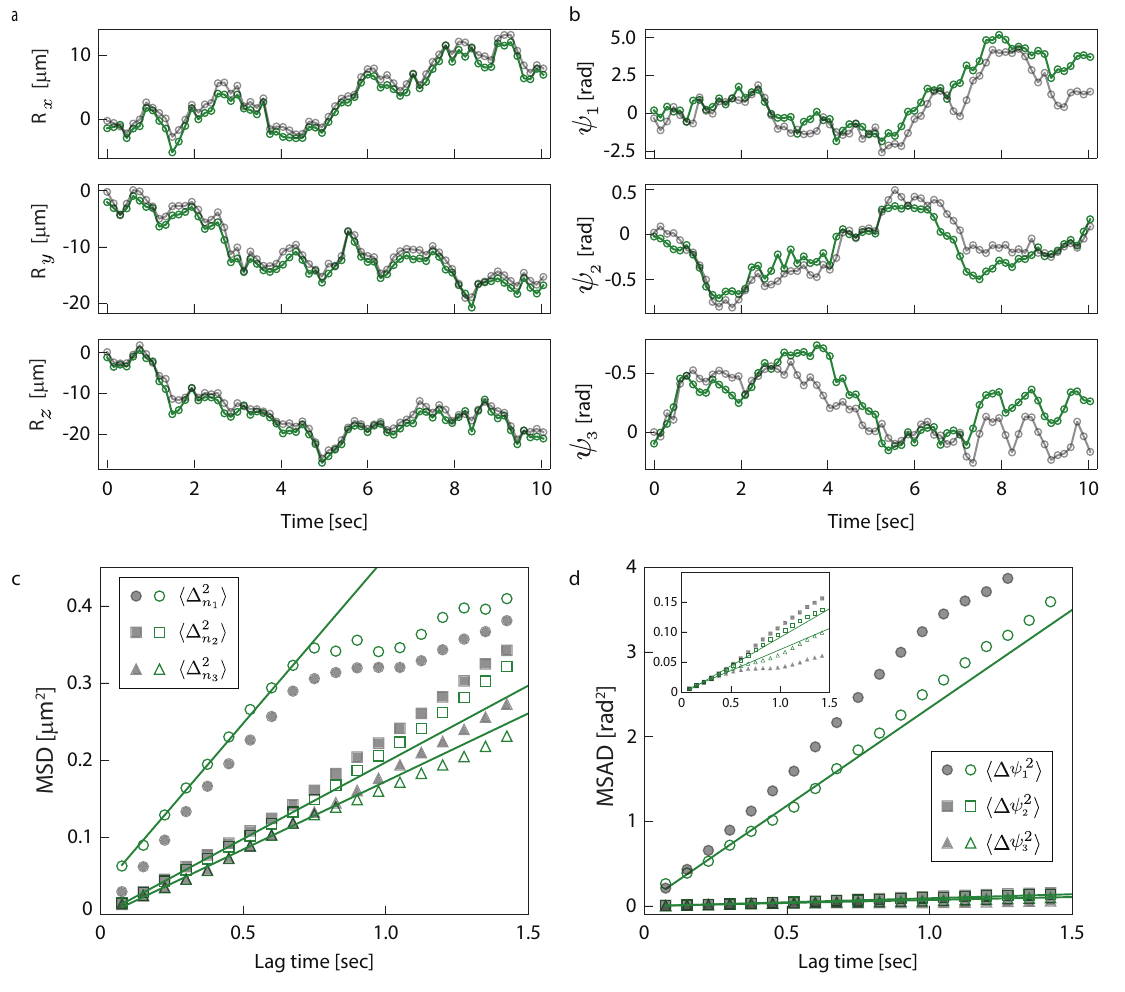}
\caption{
    \textbf{Tracking synthetic helices} 
    ({\bf a}) Tracking of the center of mass of the helix on the lab coordinates. Gray circles are the ground truth, and green circles are the computed values from the tracking algorithm. 
    ({\bf b}) Tracking of the helix orientation.
    ({\bf c}) Ground truth (grey) and computed (green) MSD along the body axes $\uvec{n}_1$ (circles), $\uvec{n}_2$ (squares), and $\uvec{n}_3$ (triangles). Solid lines are linear fits to the first ten data points. 
    ({\bf d}) Ground truth (grey) and computed (green) MSAD about the body axes $\uvec{n}_1$ (circles), $\uvec{n}_2$ (squares), and $\uvec{n}_3$ (triangles). Inset shows an enlarged version rotation about the transverse axes. 
\label{fig:SI-syn-tracking}}
\end{figure}

We created synthetic time-series images of a diffusing helix to verify our tracking algorithm. First, we created a representation of a helix in a 3D space. We determined its motion by sampling the rotational and translational displacements along the body axes by drawing from normal distributions, $\Delta_\xi \sim \mathcal{N}(0,2 D_\xi \Delta t)$. We computed the center-of-mass displacement from
\begin{equation}
\vec{R}(t + \delta t) = \vec{R}(t) + \sum_{i=1}^3 \Delta_{n_i}(\delta t) \uvec{n}_i.
\end{equation}
In this case, we neglected the propulsive coupling, which could be included by drawing from a multivariate normal distribution. We chose ground-truth diffusion coefficients similar to the experiment, $D_{n_1} = \SI[per-mode=symbol]{0.2}{\micro\meter\squared\per\second}$, $D_{n_{2,3}} = \SI[per-mode=symbol]{0.1}{\micro\meter\squared\per\second}$, $D_{\psi_1}=\SI[per-mode=symbol]{1.5}{\radian\squared\per\second}$, and $D_{\psi_{2,3}}=\SI[per-mode=symbol]{0.03}{\radian\squared\per\second}$ with the time step $\Delta t=\SI{75}{\milli \second}$.

We tracked synthetic flagella using the approach described in sections~\ref{SI:sec-segmentation}--\ref{SI:sec-msd} and found good agreement with the ground truth (Fig.~\ref{fig:SI-syn-tracking}a,b). The deviations that are present most likely come from the numerical integration required to compute the angle using eqs.(\ref{eq:3-51}--\ref{eq:3-53}).

Next, we determined the diffusion coefficients by fitting the MSDs using eq.~\ref{eq:3-54} and algorithm~\ref{alg:SI-MSD-trans}. The MSD results (Fig.~\ref{fig:SI-syn-tracking}c,d) show good agreement with the ground truth values, $D_{n_1} = \SI[per-mode=symbol]{0.22}{\micro\meter\squared\per\second}$, $D_{n_2} = \SI[per-mode=symbol]{0.1}{\micro\meter\squared\per\second}$, $D_{n_3} = \SI[per-mode=symbol]{0.09}{\micro\meter\squared\per\second}$, $D_{\psi_1} = \SI[per-mode=symbol]{1.16}{\radian\squared\per\second}$, $D_{\psi_2} = \SI[per-mode=symbol]{0.05}{\radian\squared\per\second}$, and $D_{\psi_3} = \SI[per-mode=symbol]{0.04}{\radian\squared\per\second}$. The measured translational diffusion coefficients are within \SI{10}{\percent} of the ground truth values, while the rotational coefficients have a somewhat larger deviation, but always less than \SI{50}{\percent}.

\section{Effect of tumbling on measured diffusion coefficients\label{SI:sec-simulation}}
\begin{figure*}[h!]
\centering
\includegraphics[width=1\linewidth]{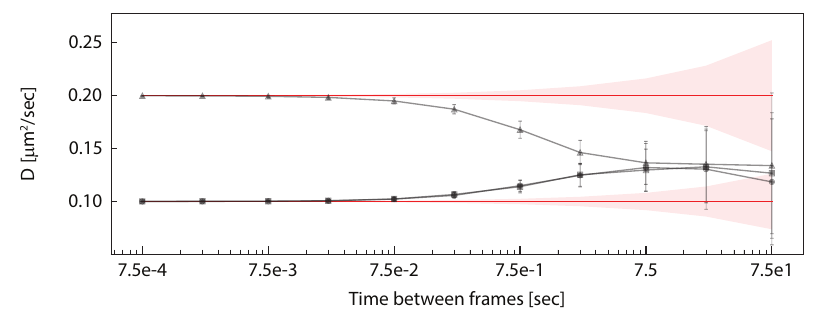}
\caption{
    \textbf{effect of tumbling on measured translational diffusion coefficients.}
    Translational diffusion coefficients versus time step size determined from simulation For the shortest time steps, the extracted values of $D_{\uvec{n}_1}$ (triangles) and $D_{\uvec{n}_{2, 3}}$ (squares and circles) match the ground truth values (red lines) while for the longest time steps they approach $D_{\text{avg}}$. Error bars are the standard deviation over trajectories. Red bands represent the statistical uncertainty in determining the diffusion coefficients from the variance of the step size.   
\label{fig:SI-simulation}}
\end{figure*}

As discussed in the main text, constant tumbling of the helix leads to the center-of-mass effectively diffusing at the mean diffusion coefficient $D_\text{avg} = (D_\parallel + 2 D_\perp)/3$ at long times. Since the translational diffusion coefficients depend on the orientation of the helix, if the helix tumbles significantly during a single volumetric image, this effectively mixes diffusion coefficients even when considering translation along the body axes. The tumbling is negligible over time step $\delta t$ if $\sqrt{D_{\psi_{2, 3}} \delta t} \ll 1$. As this quantity increases, we expect a smooth crossover between the low tumbling regime, where we can distinguish the translational diffusion coefficients, and the high tumbling regime, where we can only determine the coarse-grained diffusion coefficient $D_\text{avg}$. Although the tumbling time scales determined in the text are longer than our experiment, we investigated the effect tumbling has on our extracted body-axes diffusion coefficients by performing stochastic simulations of diffusing helices.

To determine where the experimental data falls in this crossover, we performed simulations of \num{100} trajectories of \num{3e6} time steps at a timing resolution of $\delta_t = \SI{0.75}{\milli \second}$ using ground-truth diffusion coefficients similar to the experiment, $D_{\uvec{n}_1} = \SI[per-mode=symbol]{2.4}{\micro \meter^2 \per \second}$, $D_{\uvec{n}_{2, 3}} = \SI[per-mode=symbol]{1.2}{\micro \meter^2 \per \second}$, $D_{\psi_1} = \SI[per-mode=symbol]{15}{\radian^2 \per \second}$, and $D_{\psi_{2, 3}} = \SI[per-mode=symbol]{0.3}{\radian^2 \per \second}$. As in the previous simulation, the translational and rotational steps were drawn from normal distributions at each time step. We determined the diffusion coefficients of the step-size distribution for various time-step sizes $n \delta t$ by considering every $n$th frame of the simulation. Since there is no localization error, we determined the diffusion coefficients from the variance of the step-size distribution instead of the MSD. The statistical uncertainty (standard deviation) in the diffusion coefficients extracted this way is $D \sqrt{2/(N-1)}$, determined from the variance of the sample variance of a normal distribution, where $N = \num{3e6}/n$ is the number of steps. 

The simulation results are shown in figure~\ref{fig:SI-simulation}. We found that both $D_{\uvec{n}_1}$ and $D_{\uvec{n}_{2, 3}}$ can be determined with high accuracy for time steps $\sqrt{D_\alpha n \delta t} < 0.02$, while the measured values converge to $D_\text{avg}$ for $\sqrt{D_\alpha n \delta t} > 0.5$. At the experimental value $\sqrt{D_\alpha n \delta t} \sim 0.05$, tumbling modifies the diffusion coefficients by $\sim \SI{5}{\percent}$. 

\section{Codiffusion coefficients for all coordinates\label{SI:sec-codiffusion}}
\begin{figure*}[h!]
\centering
\includegraphics[width=1\linewidth]{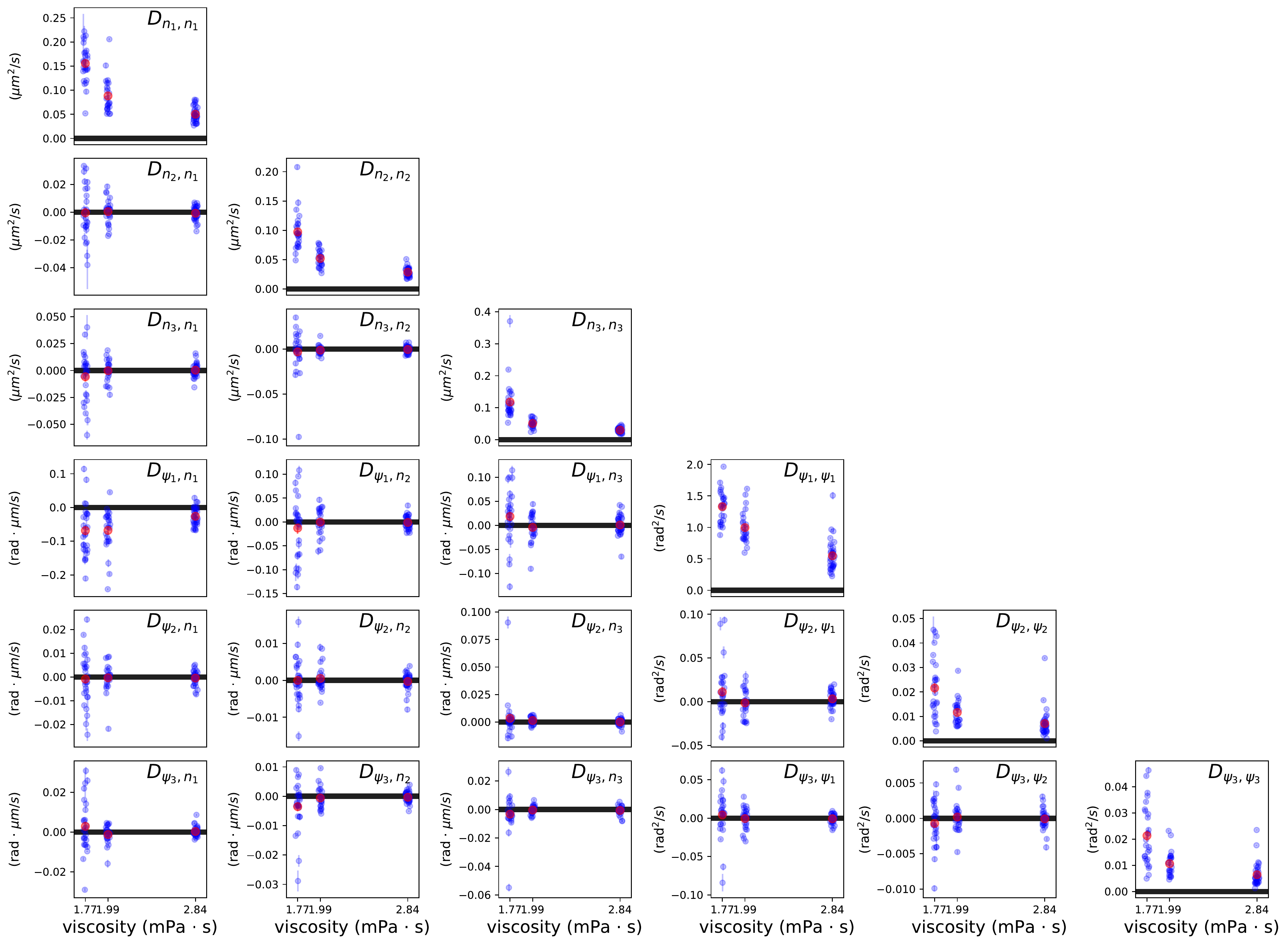}
\caption{
    {\bf Diffusion and diffusion coefficients for all coordinates.} 
    Diffusion and diffusion coefficients versus viscosity including individual flagella (blue) and mean values (red) for all $n=88$ experiments. Error bars for the individual coefficients are the standard error from the MSD fit. Error bars for the mean coefficients are sem.
\label{fig:SI-all-diffusion}}
\end{figure*}

In the main text, we focus on the diffusion coefficients $D_{n_1, \psi_1}$, motivated by the symmetry considerations of section~\ref{SI:sec-symmetry}.  However, once all the step-size data have been determined, we can compute arbitrary diffusion coefficients $D_{\xi, \eta}$ using eq.~\ref{eq:msd_fit} to test these assumptions. We show the results in figure~\ref{fig:SI-all-diffusion}. The only diffusion coefficient that is non-zero outside of uncertainty is $D_{n_1, \psi_1}$ as expected.

\section{Effect of boundary interactions on the propulsion matrix\label{SI:sec-wall}}
\begin{figure*}[h!]
\centering
\includegraphics[width=1\linewidth]{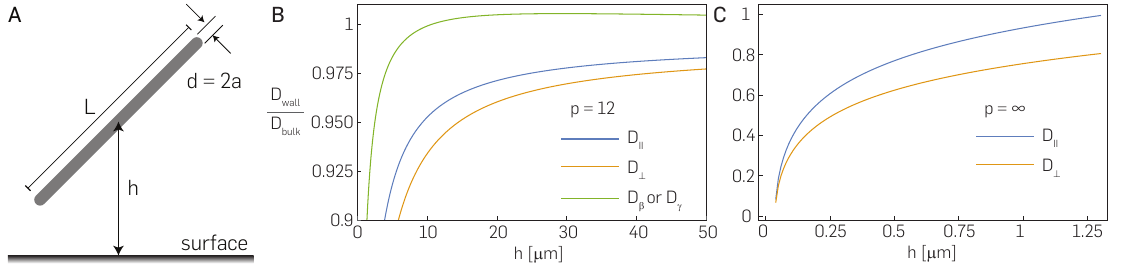}
\caption{
    {\bf Hydrodynamic interactions between a cylindrical rod and a wall.} 
    ({\bf A}) Schematic of the cylindrical rod above the coverslip surface. 
    ({\bf B}--{\bf C}) Ratio of translational and transverse rotational diffusion coefficients between the height close to the wall to the bulk for $p = 12$ and $p = \infty$ where $p = L/ 2a$.
\label{fig:2-SI-wall}}
\end{figure*}

Arbitrary bodies experience hydrodynamic interactions with boundaries that modify their propulsion matrix coefficients. As the flow field that a body produces decays strongly with distance, this effect is negligible when the body is more than a few characteristic lengths away from a boundary. In this experiment, the flagella are roughly \SI{8}{\micro \meter} long and situated \SIrange{10}{40}{\micro \meter} from the coverslip, so it is not \emph{a priori} obvious whether boundary effects are important.

Due to the complications of working with a helical geometry, we instead consider the known hydrodynamic interaction between a wall and a cylindrical rod. We expect to capture the first-order interaction between a wall and a helix. The difference between the bulk diffusion coefficients and the diffusion coefficients for a cylinder of radius $a$ and length $L$ with centroid distance $h$ from the wall (Fig.~\ref{fig:2-SI-wall}) is given by~\cite{yang2017interfacial}
\begin{align}
    \frac{D_\parallel^{T,w}}{D_\parallel^{T,b}} =&
    \frac{\left(0.9909(h/a)^3 + 0.3907(h/a)^2 - 0.1832(h/a) - 0.001815\right)}
    {\left((h/a)^3 + 2.03(h/a)^2 - 0.3874(h/a) - 0.07533\right)} \label{eq:SI-wall-1}\\
    \frac{D_\perp^{T,w}}{D_\perp^{T,b}} =&
    \frac{\left(0.9888(h/a)^3 + 0.788(h/a)^2 - 0.207(h/a) - 0.004766\right)}
    {\left((h/a)^3 + 3.195(h/a)^2 - 0.09612(h/a) - 0.1523\right)} \label{eq:SI-wall-2}\\
    \frac{D_\perp^{R,w}}{D_\perp^{R,b}} =&
    \frac{\left(0.998(h/a)^3 + 131.1(h/a)^2 + 21.25(h/a) + 0.01275\right)}
    {\left((h/a)^3 + 128.7(h/a)^2 + 121.1(h/a) + 2.897\right)}, \label{eq:SI-wall-3}
\end{align}
where the superscripts $T,w$ and $R,w$ indicate the translational and rotational diffusion coefficients close to the wall, while the superscripts $T,b$ and $R,b$ indicate the bulk diffusion coefficients. For the translation, the subscripts $\parallel$ and $\perp$ indicate translation in the direction parallel and perpendicular to the long axis of the rod. For the rotation, the subscript $\perp$ represents rotation along the transverse axes. These formulations are valid for $6 < L/2a < 30$ and the distance from the wall $h < a$ or $h > 10a$. 

To roughly match the dimensions of the helix, we chose the radius of the rod to be $a = \SI{0.25}{\micro \meter}$. At the distance $h > \SI{10}{\micro \meter}$, the wall effect on the diffusion coefficient is less than \SI{10}{\percent} as shown in figure~\ref{fig:2-SI-wall}b. We conclude that boundary effects are not important in this experiment.

\section{Determination of the viscosity inside the flow chamber\label{SI:sec-viscosity}}

\begin{figure*}[h!]
\centering
\includegraphics[width=1\linewidth]{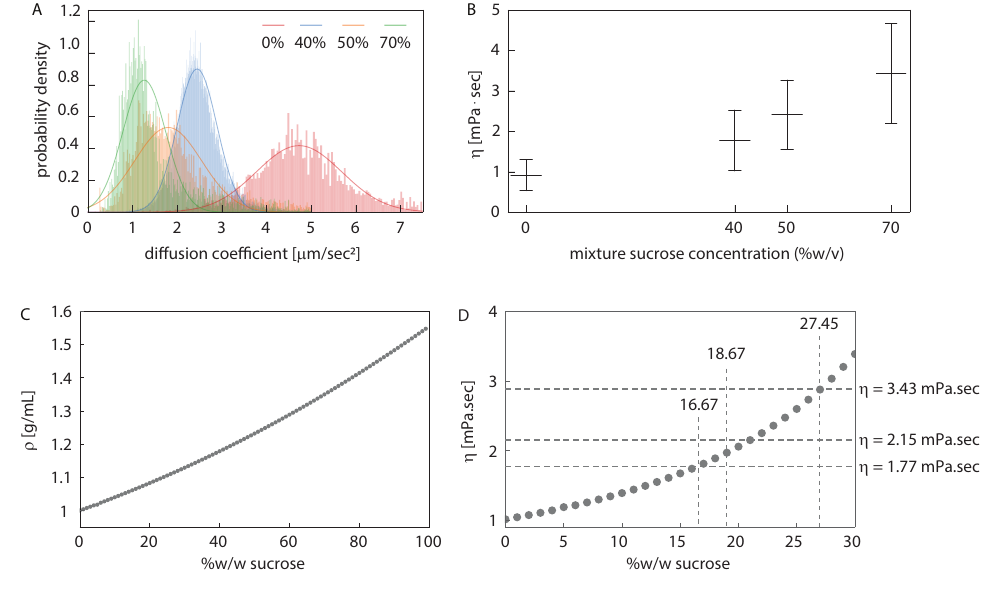}
\caption[Viscosity measurement from diffusing beads]{
    {\bf Viscosity measurement using diffusing beads.} 
    ({\bf A}) Histograms of diffusion coefficients for beads suspended in indicated sucrose concentration \%(w/w) at \SI{25}{\degreeCelsius}. Solid lines are best gaussian fit of the distributions.  
    ({\bf B}) Estimated values of viscosity at various sucrose concentrations from Einstein-Stokes relation
    ({\bf C}) Sucrose density as a function of sucrose concentration \%(w/w) determined experimentally at \SI{20}{\degreeCelsius}~\cite{asadi2006beet}. 
    ({\bf D}) Theoretical viscosity at \SI{20}{\degreeCelsius} versus sucrose concentration. Highlighted viscosity values are the numbers obtained from the diffusing bead measurements.
\label{fig:SI-vis-diff}}
\end{figure*}

Due to the preparation of the chamber slides with BSA, the sucrose solution present on the slide during the experiment is more dilute than the solution that was initially prepared. Since the final dilution is unknown, the viscosity cannot be calculated exactly. Instead, we determined the viscosity for each sucrose dilution by measuring diffusion coefficients of \SI{99}{\nano\meter} diameter Tetraspek fluorescent beads (Invitrogen, T7279) in an auxiliary sample using an Oxford Nanoimager operating with HILO illumination using a \SI{532}{\nano \meter} laser incident at \ang{25} relative to the coverslip. Single-particle tracking and diffusion coefficient analysis were performed using NimOS software. Figure~\ref{fig:SI-vis-diff}a shows histograms of diffusion coefficients of beads measured at various concentrations of sucrose mixtures. The viscosity was then inferred from the Stokes-Einstein relationship, $\eta = \frac{k_B T}{6 \pi D_o r}$ where $T$ is the absolute temperature, $D_o$ is the measured diffusion coefficient, and $r$ is the bead radius. Figure~\ref{fig:SI-vis-diff}b shows the viscosity calibration. 

The viscosity and the sucrose molar volume at \SI{20}{\degreeCelsius} are related by~\cite{soesanto1981volumetric},
\begin{equation}
    \eta = \SI{6.31e-3}{\milli \pascal \second} \times \exp \left( \SI[per-mode=symbol]{282}{\mole \per \liter} \times V_m \right), \label{eq:SI-vis-1}
\end{equation}
where $V_m$ is the molar volume in [L mol$^{-1}$]. The molar volume $V_m$ is related to the concentration in \%w/w by
\begin{eqnarray}
    V_m &=& \frac{x_{s} MW_{s} + (1-x_{s}) MW_{w}}{\rho_\text{mix}}\label{eq:SI-vis-2}\\
    x_s &=& \frac{\frac{1}{MW_s}}{\frac{1}{MW_s} + \frac{1-\%(w/w)}{\%(w/w)} \frac{1}{MW_w}}. \label{eq:SI-vis-7}
\end{eqnarray}
where $x_{s}$, $MW_{s}$, and $MW_{w}$ are the molar fraction of sucrose ($\text{C}_{12}\text{H}_{22}\text{O}_{11}$), the molecular weight of sucrose, and the molecular weight of water, respectively. The denominator $\rho_\text{mix}$ is the density of the sucrose and water mixture.

Combining eqs.~\ref{eq:SI-vis-2} and \ref{eq:SI-vis-7} with the known concentrations and the measured values of $\rho_\text{mix}$ from~\cite{asadi2006beet}, we found that the viscosities and sucrose concentrations for the three dilutions used in this work were $\eta = \num{1.77}$, \num{2.15}, and \SI{3.43}{\milli \pascal \second} and  \%(w/w) = \SI{16.67}{\percent}, \SI{18.96}{\percent} and \SI{27.01}{\percent} (Fig.~\ref{fig:SI-vis-diff}c,d).

\section{Diffusion of microtubules}\label{SI:sec-microtubules}

\begin{figure*}[h!]
	\centering
	\includegraphics[width=1\linewidth]{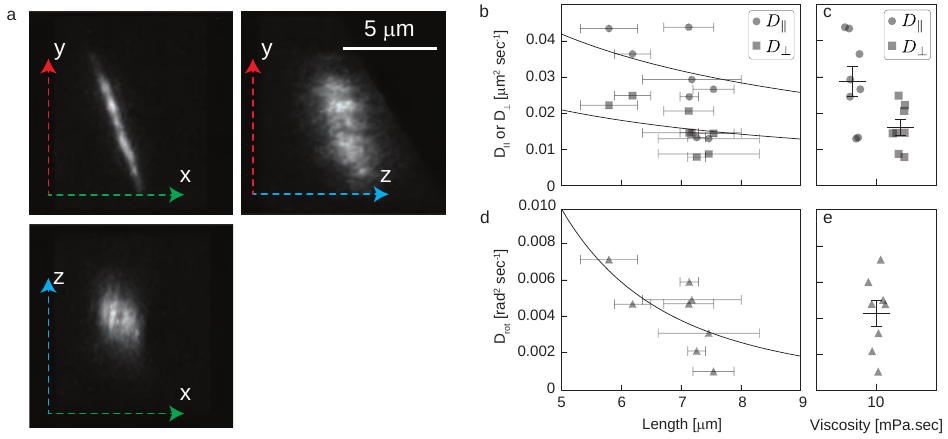}
    \caption[Diffusion coefficients of microtubules]{{\bf Diffusion coefficients of microtubules measured from the oblique plane light sheet microscope.}
    ({\bf A}) Orthographic projection of a 3D microtubule after deconvolution and deskewing. 
    ({\bf B}) Translational diffusion coefficients versus microtubule lengths along the longitudinal (blue) and transverse (orange) axes. The solid line represents the theoretical prediction.
    ({\bf C}) Translational diffusion coefficients along the longitudinal (blue) and transverse (orange) axis. Error bars represent standard deviation.
    ({\bf D}) Rotational diffusion coefficients about the transverse axis.
    ({\bf E}) Rotational diffusion coefficients along the transverse axis. Error bars represent standard deviation.
    \label{fig:SI-microtubules}}
\end{figure*}

To further validate our experimental and data analysis frameworks, we performed a control experiment using microtubules, which approximate straight and rigid rods. The diffusing microtubules, with an average length of \SI{6.96(22)}{\micro\meter} ($n = 6$) were immersed in a fluid medium with a viscosity of \SI{\sim 10}{\milli\pascal\second}. Compared to flagella, we used a higher viscosity solution to slow the motion of the microtubules. We tracked the microtubules using the same approach as for the flagella, although in this case, due to symmetry, it is not possible to resolve the rotation about $\uvec{n}_1$. From the tracking, we computed MSDs and MSADs using Alg.\ref{alg:SI-MSD-trans} (SI section \ref{SI:sec-data-analysis}) and fit the first \num{10} data points to determine the diffusion coefficients.

The translational diffusion coefficients along the longitudinal axis and transverse axes are \SI[per-mode=symbol]{0.028(004)}{\micro\meter\squared\per\second} and \SI[per-mode=symbol]{0.015(002)}{\micro\meter\squared\per\second}. The ratio between these values is \num{1.78(09)}, whereas the ratio is expected to be exactly \num{2} for a cylinder~\cite{cox1970motion}. Potential errors may be related to the tumbling rate, or microtubules are not perfectly rigid. Indeed, the persistence length of GMPCPP-stabilized microtubules is \SI{1.8\pm 0.5}{\milli\meter}~\cite{hawkins2013mechanical}.

The measured microtubule diffusion coefficients are compared with the theoretical values for a cylinder of length $L$ and diameter $d$, which are~\cite{Broersma1960, li2004diffusion}
\begin{eqnarray}
    D_\parallel &=& \frac{k_B T\left[{\ln \left(L/d\right) + \gamma_\parallel}\right]}{2\pi\eta L} \label{eq:dparallel_rod}\\
    D_\perp &=& \frac{k_B T\left[{\ln \left(L/d \right) + \gamma_\perp}\right]}{4\pi\eta L} \label{eq:dperp_rod}\\
    D_r &=& \frac{3 k_B T\left[{\ln \left(L/d \right) + \gamma_r}\right]}{2\pi\eta L^3}.  \label{eq:drot_rod}
\end{eqnarray}
For $L/d = \infty$, $\gamma_\parallel=-0.114$, $\gamma_\perp = 0.886$, and $\gamma_r = -0.447$~\cite{broersma1981viscous}.

\section{Non-dimensional propulsion matrix}\label{SI:sec-nondim}
The propulsion matrix coefficients depend only on the viscosity and helical geometry, and hence can be parameterized by the helical radius $R$, length $L$, filament radius $a$, and helical pitch $\lambda$. To allow comparison of our work with prior results, we define non-dimensionalized propulsion matrix coefficients $A^* \equiv A / \eta L$, $B^* \equiv B / \eta LR$, and $D^* \equiv D / \eta LR^2$. While the non-dimensional propulsion matrix removes all dependence on viscosity and first-order dependence on the flagellar length and helical diameter based on RFT, it does not remove the dependence on the helical pitch (equivalently $R / \lambda$) or $\lambda /a$.

To determine the non-dimensional propulsion matrix elements from our measured diffusion coefficients, we relied on experimentally measured geometric parameters of \emph{E. coli} flagella. From image analysis of the $n = 88$ flagella, we find $l = \SI{7.7(14)}{\micro\meter}$ and $\lambda = \SI{2.5(1)}{\micro \meter}$, values that are consistent with previous measurements~\cite{turner2000real,Darnton2007}. Unlike in the viscosity experiments ($n=77$) we consider flagella of all lengths here. As the helical radius is on the order of the diffraction limit, we rely on the previously reported value $R = \SI{0.25}{\micro \meter}$, which implies that the helical pitch angle, $\theta = \tan^{-1}\left( 2\pi R / \lambda \right)$, is $\theta = \ang{32}$. As the filament radius is much smaller than the diffraction limit in our setup, we rely on the value obtained by x-ray diffraction, $a = \SI{0.01}{\micro \meter}$~\cite{Namba1989,Samatey2001}.

In Fig.~4d of the main text, we compare the non-dimensionalized propulsion matrix elements obtained from Brownian motion versus conventional hydrodynamics setup for helices with $\theta \sim \ang{32}$. We consider three experimental setups and two theory techniques. Optical tweezer values~\cite{chattopadhyay2006swimming} have been non-dimensionalized using their parameter estimates $R = \SI{0.21}{\micro \meter}$, $L = \SI{6.5}{\micro \meter}$, and $\theta = \SI{41}{\degree}$. Load cell values~\cite{rodenborn2013propulsion} were taken from their figure 2 using $\theta = \ang{32}$. Gravity values from~\cite{purcell1997efficiency} were taken from the second entry in Table 1, with $\theta = \ang{39}$. Johnson SBT and regularized Stokeslet values were obtained using the code developed in~\cite{rodenborn2013propulsion} with the flagella parameters given above. 

The mm-scale helix experiments agree and show reasonably close agreement with slender body theory (SBT) and regularized Stokeslet calculations~\cite{Johnson1980, Cortez2001, rodenborn2013propulsion}. Some of the remaining deviation between the theory and these experiments comes from using $\lambda / a \sim \num{200}$ which is accurate for the flagella but somewhat too small for the wire helices $\lambda / a \sim \num{40}$~\cite{rodenborn2013propulsion}.

\end{document}